\documentclass[letter]{aa} 
\usepackage{graphicx}
\usepackage{txfonts}
\usepackage{natbib}     
\usepackage{xcolor} 
\usepackage{threeparttable} 

\bibpunct{(}{)}{;}{a}{}{,} 

\begin{document}

   \title{Detection of CO emission lines in the dayside atmospheres of WASP-33b and WASP-189b with GIANO}
 
   \author{F. Yan\inst{1}
                  \and
          E. Pall\'e\inst{2,3}
                  \and
                  A. Reiners\inst{1}
                  \and
                  N. Casasayas-Barris\inst{4}
                  \and
          D. Cont\inst{1}
                  \and
          M.~Stangret\inst{2,3}
                  \and
          L.~Nortmann\inst{1}
          \and
          P.~Molli\`ere\inst{5}
          \and
                 Th.~Henning\inst{5}
          \and
         G.~Chen\inst{6,7}
          \and
          K. Molaverdikhani\inst{8,9,5}
                }

  \institute{Institut f\"ur Astrophysik, Georg-August-Universit\"at, Friedrich-Hund-Platz 1, D-37077 G\"ottingen, Germany\\
        \email{fei.yan@uni-goettingen.de}
        \and
        Instituto de Astrof{\'i}sica de Canarias (IAC), Calle V{\'i}a Lactea s/n, E-38200 La Laguna, Tenerife, Spain
\and
Departamento de Astrof{\'i}sica, Universidad de La Laguna, 38026  La Laguna, Tenerife, Spain
\and
Leiden Observatory, Leiden University, Postbus 9513, 2300 RA, Leiden, The Netherlands
\and
Max-Planck-Institut f{\"u}r Astronomie, K{\"o}nigstuhl 17, 69117 Heidelberg, Germany
\and
Key Laboratory of Planetary Sciences, Purple Mountain Observatory, Chinese Academy of Sciences, Nanjing 210023, China
\and
CAS Center for Excellence in Comparative Planetology, Hefei 230026, China
\and
Universit\"ats-Sternwarte, Ludwig-Maximilians-Universit\"at München, Scheinerstrasse 1, D-81679 München, Germany
\and
Exzellenzcluster Origins, Boltzmannstraße 2, 85748 Garching, Germany
}
        \date{Received March 9, 2022; accepted April 20, 2022}


  \abstract
 {Ultra-hot Jupiters (UHJs) are expected to possess temperature inversion layers in their dayside atmospheres. Recent thermal emission observations  have discovered several atomic and molecular species along with temperature inversions in UHJs. We observed the thermal emission spectra of two UHJs (WASP-33b and WASP-189b) with the GIANO-B high-resolution near-infrared spectrograph. Using the cross-correlation technique, we detected carbon monoxide (CO) in the dayside atmospheres of both planets. 
The detected CO lines are in emission, which agrees with previous discoveries of iron emission lines and temperature inversions in the two planets. 
This is the first detection of CO lines in emission with high-resolution spectroscopy. Further retrieval work combining the CO lines with other spectral features will enable a comprehensive understanding of the atmospheric properties such as temperature structures and C/O ratios. The detected CO and iron emission lines of WASP-189b have redshifted radial velocities of several km\,s$^{-1}$, which likely originate from a dayside to nightside wind in its atmosphere. Such a redshifted velocity has not been detected for the emission lines of WASP-33b, suggesting that the atmospheric circulation patterns of the two UHJs may be different.
 }

   \maketitle

%

\section{Introduction}
Characterizing the atmosphere of ultra-hot Jupiters (UHJs) has become a fast expanding subject in the field of exoplanet science. Ground-based high-resolution spectroscopy, in particular, plays an important role in discovering chemical species in the atmosphere of UHJs. For example, many species have been detected via high-resolution transmission spectroscopy in the hottest known exoplanet -- KELT-9b, including hydrogen Balmer lines, \ion{Fe}{i}, \ion{Fe}{ii}, \ion{Ti}{ii}, \ion{Ca}{ii},  \ion{Mg}{i}, and \ion{O}{i} \citep{Yan2018, Hoeijmakers2018, Hoeijmakers2019, Yan2019, Cauley2019, Turner2020, Wyttenbach2020, Borsa2021, PaiAsnodkar2022}. 
Various chemical species have also been discovered in other UHJs such as KELT-20b/MASCARA-2b \citep{Casasayas-Barris2018, Casasayas-Barris2019, Stangret2020, Hoeijmakers2020, Rainer2021}, WASP-76b \citep{Seidel2019, Zak2019, Ehrenreich2020, Tabernero2021, Casasayas-Barris2021, Deibert2021, Landman2021, Kesseli2022}, WASP-121b \citep{Sing2019, Bourrier2020, Cabot2020,Ben-Yami2020, Hoeijmakers2020-W121, Borsa2021-W121}, TOI-1518b \citep{Cabot2021}, and HAT-P-70b \citep{Bello-Arufe2022}.

Thermal emission spectroscopy has also been used in characterizing the dayside atmosphere of UHJs. For example, atomic \ion{Fe}{i} emission lines have been detected in KELT-9b \citep{Pino2020,Kasper2021}, WASP-189b \citep{Yan2020}, WASP-33b \citep{Nugroho2020W33, Cont2021}, and KELT-20b \citep{Yan2022}. We note that OH and TiO emission lines were found in WASP-33b \citep{Nugroho2021,Cont2021}. \cite{Cont2022} have recently reported the detection of \ion{Si}{i} in WASP-33b and KELT-20b. The spectral lines of these detected species are all in emission, which unambiguously indicates the presence of temperature inversion layers on the dayside of these UHJs. These observational results agree with theoretical simulations, which predict the existence of temperature inversions in UHJs \citep[e.g.,][]{Lothringer2018, Kitzmann2018, Helling2019, Fossati2021-NLTE}.

In addition to these high-resolution studies, space-based low-resolution emission spectroscopy has also been widely used in studying temperature inversions. For example, strong $\mathrm{H_2O}$ emission features have been found in WASP-121b \citep{Evans2017} and KELT-20b \citep{Fu2022} with the Hubble Space Telescope. Evidence of carbon monoxide (CO) emission has been inferred from the Spitzer photometric observations at 3.6 $\mathrm{\mu}$m in several UHJs (e.g., WASP-103, WASP-18b, and KELT-20b) \citep{Sheppard2017, Kreidberg2018}. 

Here, we present the high-resolution detection of CO emission lines in the dayside atmospheres of two UHJs (\object{WASP-33b} and \object{WASP-189b}) with the GIANO-B spectrograph. This is the first detection of CO lines with an emission line profile since previous high-resolution CO lines were only detected in absorption \citep[e.g.,][]{Snellen2010, Brogi2012, Konopacky2013, Line2021, Giacobbe2021}.
We notice that, during the refereeing process of this manuscript, \cite{vanSluijs-2022} also reported the detection of CO emission lines in WASP-33b.

The atmospheres of the two UHJs presented in this work were observed with other instruments before. For WASP-33b, hydrogen Balmer lines and \ion{Ca}{ii} have been detected in its transmission spectrum \citep{Yan2019, Yan2021-W33} and emission lines of \ion{Fe}{i}, OH, TiO, and \ion{Si}{i} have been discovered on WASP-33b's dayside. For WASP-189b, \ion{Fe}{i} emission lines have been detected in its dayside atmosphere \citep{Yan2020}, and various chemical species (e.g., \ion{Fe}{i}, \ion{Fe}{ii}, \ion{Ti}{i}, \ion{Cr}{i}, and TiO) have been detected in its transmission spectrum \citep{Stangret2021, Prinoth2022}. The detected transmission lines of these species have different radial velocities (RVs), indicating a three-dimensional thermo-chemical stratification at the terminator of WASP-189b \citep{Prinoth2022}.

The paper is organized as follows. We describe the observations and data reduction in Section 2 and the methods in Section 3. In Section 4, we present the results and discussions. The conclusions are provided in Section 5.

%

\section{Observations and data reduction}
\begin{figure}
   \centering
   \includegraphics[width=0.45\textwidth]{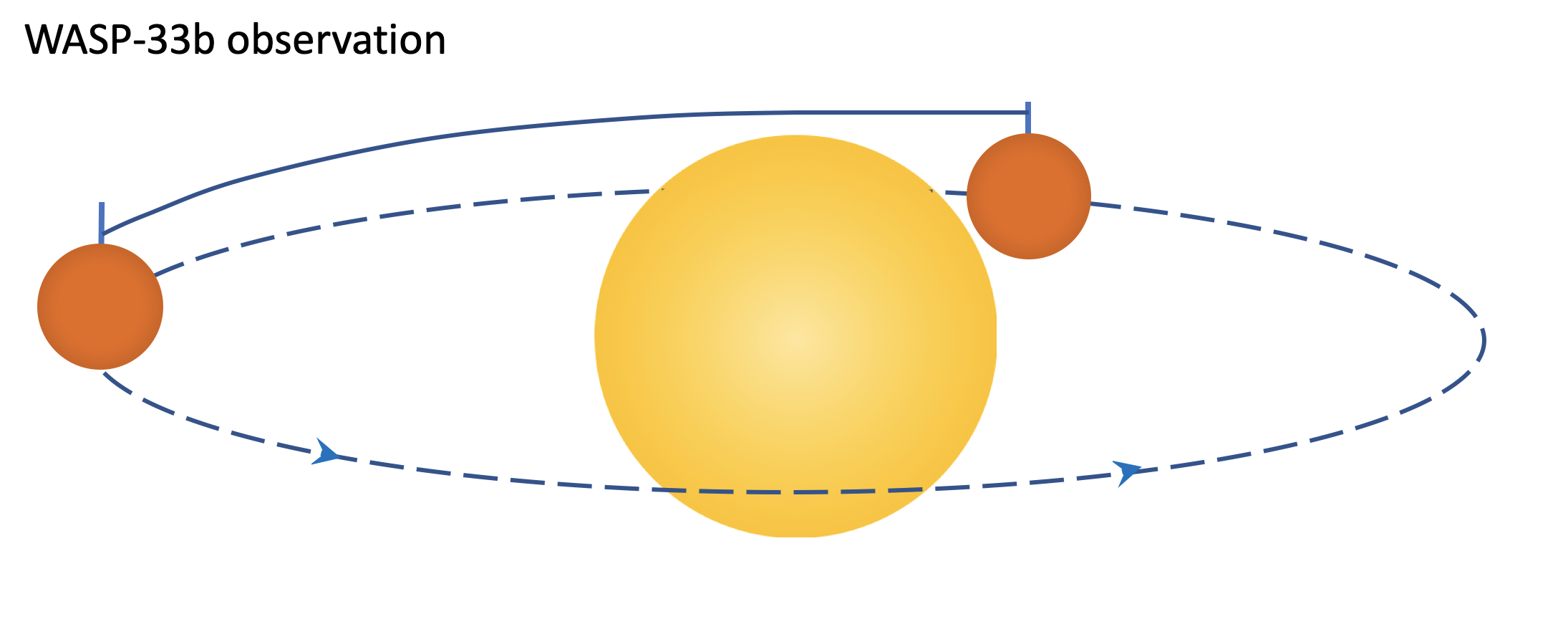}
   \includegraphics[width=0.45\textwidth]{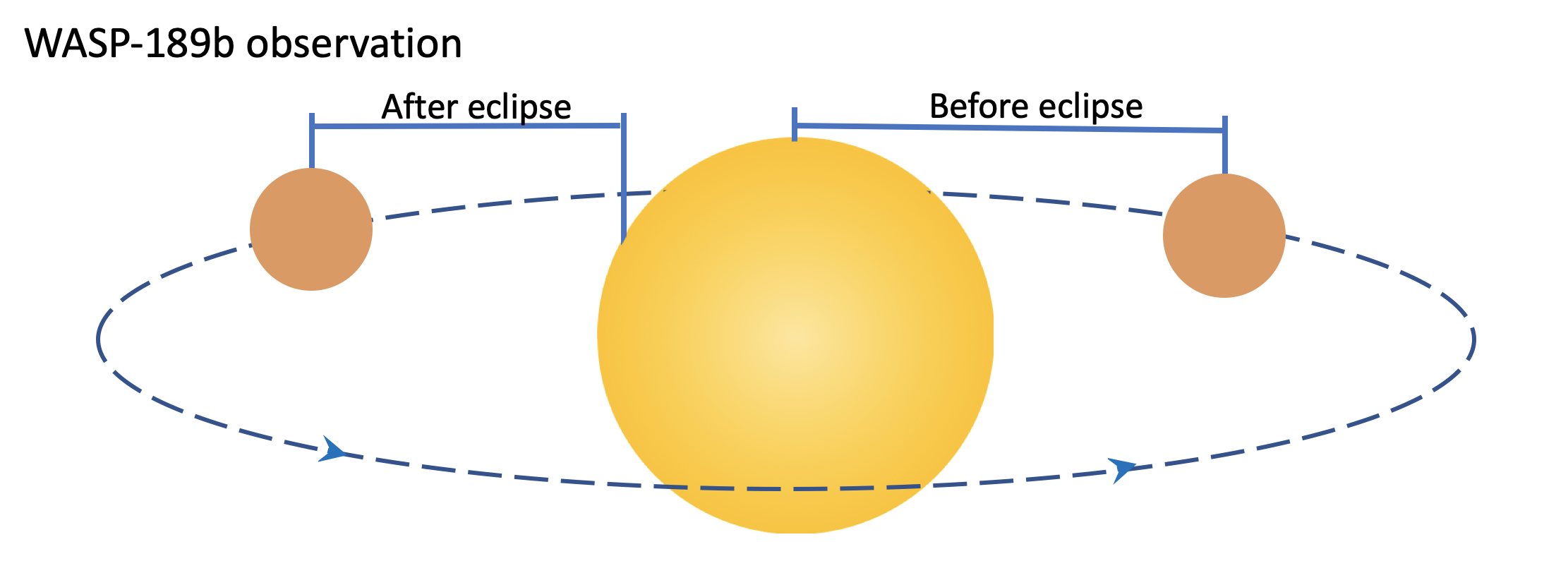}
      \caption{Schematic of the orbital phase coverage of the GIANO-B observations. }
         \label{orbit}
   \end{figure}

We observed WASP-33b on 15 October 2020 and WASP-189b on 13 April 2019 and 29 April 2019 with the GIARPS instrument \citep{Claudi2016} mounted on the Telescopio Nazionale \textit{Galileo}.
To observe the dayside hemispheres, we performed the observations at planetary orbital phases near the secondary eclipse (Fig.~\ref{orbit}). The detailed observation logs are presented in Table \ref{obs_log}.
The GIARPS instrument is the combination of the HARPS-N  and GIANO-B spectrograph, which allows for simultaneous observations in the visible and near-infrared (NIR) wavelengths. The HARPS-N data have been used to study the \ion{Fe}{i} emission lines in WASP-189b \citep{Yan2020} and WASP-33b \citep{Cont2021}. In this work, we use only the GIANO-B data to search for CO lines. GIANO-B is a NIR high-resolution spectrograph (\textit{R} $\sim$ 50\,000) that consists of 50 echelle orders with a wavelength coverage of 0.95--2.45\, $\mathrm{\mu m}$ \citep{Claudi2017}. The observations were carried out in the ABAB nodding mode, which enables the subtraction of sky background and detector noise using the AB pairs of the spectra. 

The raw spectra were reduced with the \texttt{GOFIO} pipeline \citep{Rainer2018}. The pipeline reduced spectra consist of two data sequences -- one in nodding position A and the other in nodding position B. The A and B spectra were treated separately in the following reduction procedures. 
We normalized the spectra order-by-order using a polynomial fit. Because each spectral order spans two detectors with a flux jump in the middle, we divided each order into two sections and normalized them separately.

The wavelength solution of the GIANO-B spectrograph is known to be unstable between consecutive exposures \citep{Brogi2018}. Therefore, we performed additional wavelength corrections following the method in \cite{Guilluy2019} and \cite{Giacobbe2021}. Since the CO lines are located at the first two spectral orders of the spectrograph (i.e., the reddest two orders), we only performed the correction for these two orders. For each data set, we first aligned each observed spectrum to the averaged spectrum using the telluric lines in order to correct for the spectral drift. The calculated drift of the two spectral orders was small for the entire data set (< 0.2 pixel), which agrees with previous work \citep{Brogi2018}.
We then refined the wavelength solutions of the pipeline by matching the observed telluric lines to a theoretical telluric template calculated using the \texttt{Molecfit} tool \citep{Smette2015}. The precision of the refined wavelength solution is around 1 km\,s$^{-1}$, which was estimated by computing the standard deviation of the calibration residuals.

%
\begin{table*}
\caption{Observation logs.}             
\label{obs_log}      
\centering                          
\begin{threeparttable}
        \begin{tabular}{l c c c c c }        
        \hline\hline    \noalign{\smallskip}               
        Target &        Date & Airmass change & Phase coverage & Exposure time & $N_\mathrm{spectra}$ (total/out-of-eclipse) \\     
        \hline  \noalign{\smallskip}                  
  WASP-33b & 2020-10-15  & 2.02 -- 1.01 -- 1.64 &  0.43 -- 0.74  & 200 s & 122/88 \\ 
  \hline  \noalign{\smallskip}                                
  WASP-189b & 2019-04-13         & 1.79 -- 1.18 -- 1.59 &  0.53 -- 0.62  & 60 s & 170/168 \\ 
  WASP-189b & 2019-04-29         & 2.08 -- 1.18 -- 2.19 & 0.38 -- 0.50 & 60 s & 210/152 \\                       
\hline 
        \end{tabular}
\end{threeparttable}      
\end{table*}

\section{Methods}
To search for CO signals in the dayside spectra, we applied the cross-correlation technique by cross-correlating the observed spectra with modeled spectral templates.
We first performed the \texttt{SYSREM} algorithm \citep{Tamuz2005, Birkby2013} to remove the telluric lines and stellar lines at the observer's rest frame. The observed spectra are dominated by telluric absorption lines because the A-type host stars have few spectral lines at these wavelengths. We masked the strong telluric lines by discarding the pixels with low signal-to-noise ratios (S/Ns). We used the S/N values provided by the pipeline and calculated the mean S/N for each pixel. In this way, we discarded $10\%$--$20\%$ of the pixels for each data set. We then applied the \texttt{SYSREM} algorithm to each data set in the same way as in \cite{Gibson2020} and \cite{Yan2022}. 
We tested the \texttt{SYSREM} with iterations from 1 to 15. 
The final CO detection significance reached an asymptote after the initial few \texttt{SYSREM} iterations and we chose the number that yields the maximum detection significance (Fig.~\ref{App-sysrem}).
The telluric and stellar lines' removed spectra were then shifted to the stellar rest frame by correcting the observer's barycentric RV and the stellar systemic RV.

\begin{figure}
   \centering
   \includegraphics[width=0.45\textwidth]{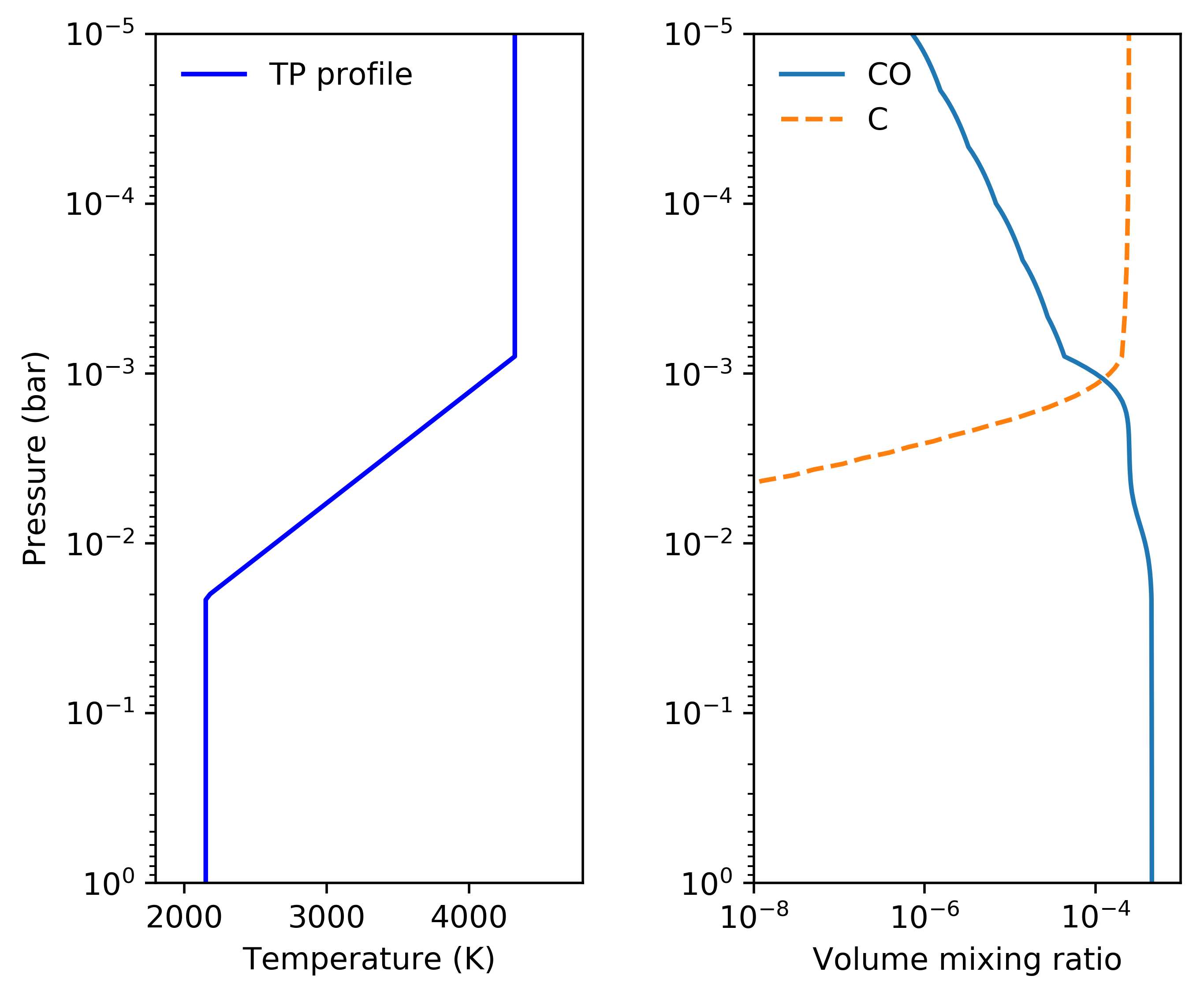}
      \caption{Adopted $T-P$ profile (left) and CO mixing ratio (right) for the calculation of the spectral template.}
         \label{TP}
   \end{figure}

We modeled the thermal emission spectrum using the \texttt{petitRADTRANS} tool \citep{Molliere2019}. The CO opacities (up to 5000\,K) were calculated using the line list from \cite{Li2015} with the \texttt{ExoCross} tool \citep{Yurchenko2018}.
We adopted the temperature-pressure ($T$-$P$) profile that was retrieved from \ion{Fe}{i} emission lines of WASP-189b \citep{Yan2020}. This $T$-$P$ profile consists of a strong temperature inversion (Fig.~\ref{TP}).
Then we calculated the CO mixing ratio using the \texttt{easyCHEM} code \citep{Molliere2015} under the chemical equilibrium and solar abundance assumptions.
According to the simulation (right panel in Fig.~\ref{TP}), CO is abundant at lower altitudes and it begins to be thermally dissociated in the temperature inversion layer. At very high altitudes, CO is largely dissociated into atomic carbon and oxygen. The planetary dayside spectra ($F_\text{p}$) of WASP-189b and WASP-33b were calculated using the parameters in Tables \ref{paras_W33} and \ref{paras_W189}. 
The observed spectrum also contains the flux from the star; therefore, the model spectrum should be $F_\text{s}$ + $F_\text{p}$, where we assumed the stellar flux $F_\text{s}$ to be a blackbody spectrum. Since the observed spectrum has been normalized, we expressed the template spectrum as 1 + $F_\text{p}$/$F_\text{s}$ (Fig.~\ref{template}). 
We then calculated a grid of the template spectrum by shifting the modeled spectrum from --\,500 km\,s$^{-1}$ to +\,500 km\,s$^{-1}$ in 1 km\,s$^{-1}$ steps.
The template spectrum was further normalized, convolved with a Gaussian function corresponding to the instrumental resolution (50\,000), and sampled to the same wavelength grid of the observed spectra.
Before the cross-correlation, we filtered both the observed and template spectra with a Gaussian high-pass filter of 15 points. In this way, we removed broadband spectral features in the spectra and, therefore, both the data and the template have values around one.

The cross-correlation procedure was performed by calculating the weighted cross-correlation function (CCF) \citep{Gibson2020} using the equation
\begin{equation}
\mathrm{CCF} = \sum_{i} \frac{R_i \, m_i}{\sigma_i^2},
\end{equation}
where $R_i$ is the observed spectrum, $m_i$ is the template spectrum, and $\sigma_i$ is the noise of the observed data at wavelength point $i$. We took the noise values provided by the instrumental pipeline and considered error propagation during the data reduction. 
The weighted CCF method normally yields a better final detection significance because the low-S/N pixels (e.g., inside the telluric lines) and the low-flux spectra are assigned with low weights.

\section{Results and discussions}
\subsection{Detection of the CO lines}
\begin{figure}
   \centering
   \includegraphics[width=0.45\textwidth]{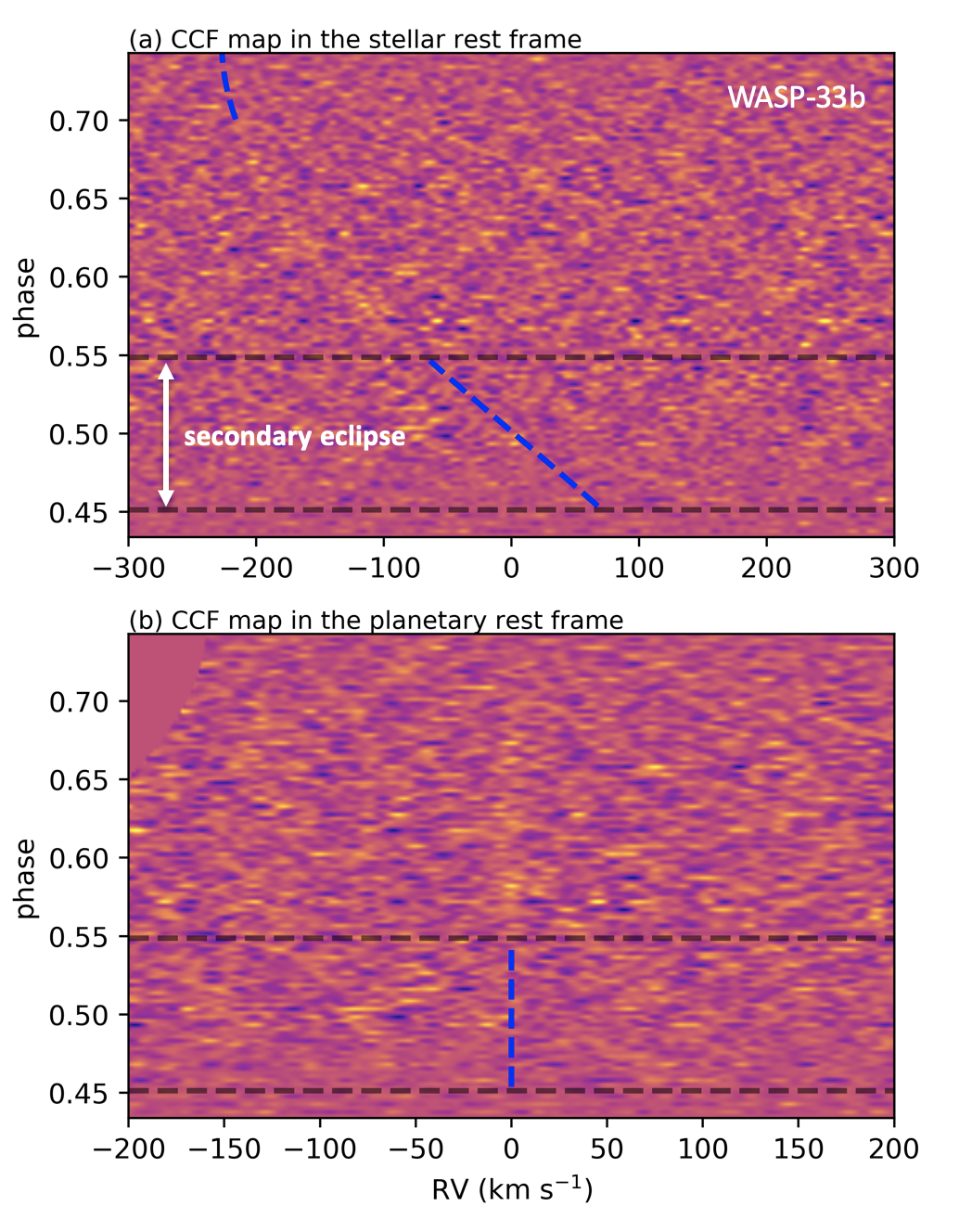}
      \caption{CCF map for WASP-33b in the stellar rest frame (upper panel) and the planetary rest frame (lower panel). The blue dashed line indicates the RV of the planetary orbital motion.}
         \label{CCF-W33}
   \end{figure}

\begin{figure}
   \centering
   \includegraphics[width=0.45\textwidth]{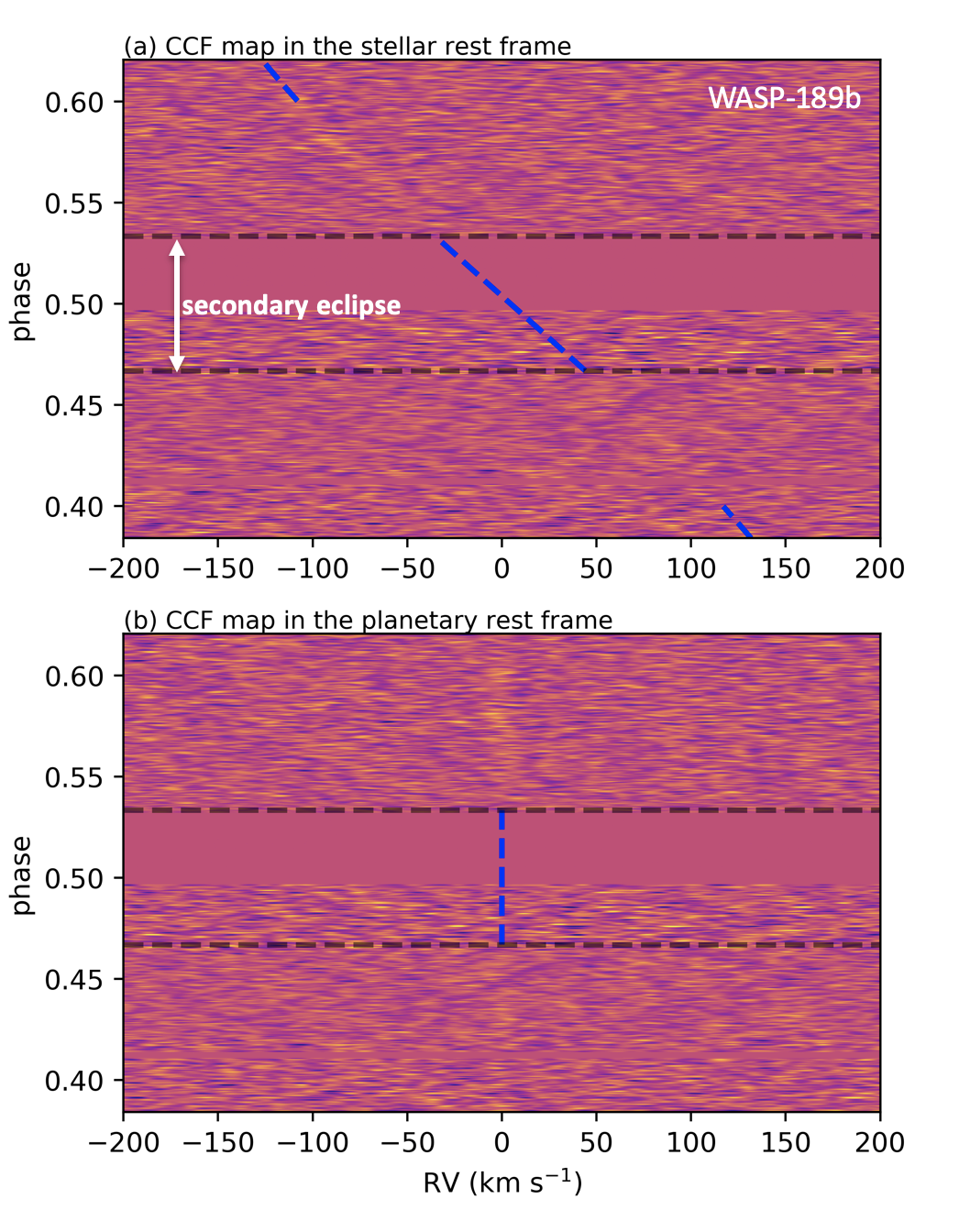}
      \caption{Same as Fig.~\ref{CCF-W33}, but for WASP-189b.}
         \label{CCF-W189}
   \end{figure}

The upper panels of Figs.~\ref{CCF-W33} and \ref{CCF-W189} present the calculated CCFs of all the observed spectra in the stellar rest frame. Although the CCFs for position A and position B were calculated separately, we merged the A and B sequences in the figures. The trails of the CO emission signal are visible for both WASP-33b and WASP-189b in the CCF maps and their RVs are aligned with the planetary orbital motion (blue dashed lines). The lower panels of Figs.~\ref{CCF-W33} and \ref{CCF-W189} show the CCFs in the planetary rest frame by correcting the planetary orbital RV.

\begin{figure}
   \centering
   \includegraphics[width=0.45\textwidth]{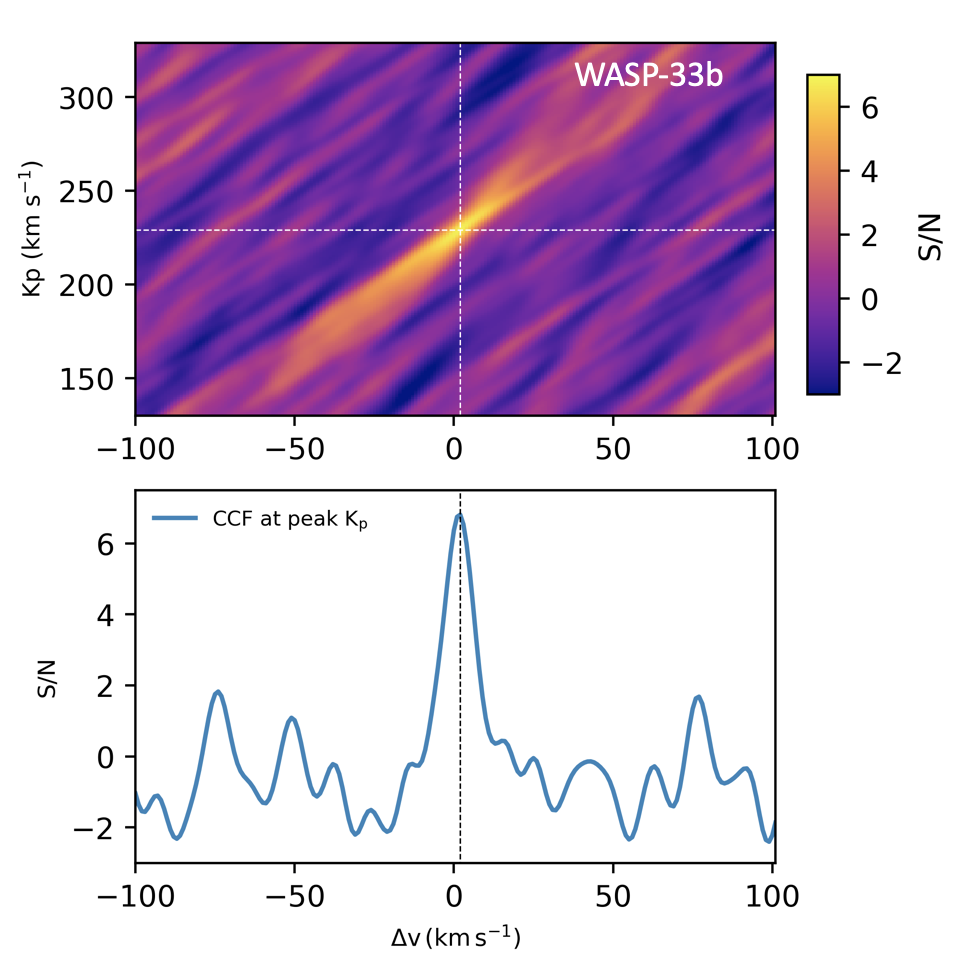}
      \caption{$K_\mathrm{p}$ map and the CCF at the maximum S/N location for WASP-33b. The white dashed lines indicate the location of the maximum S/N.}
         \label{Kp-W33}
   \end{figure}

\begin{figure}
   \centering
   \includegraphics[width=0.45\textwidth]{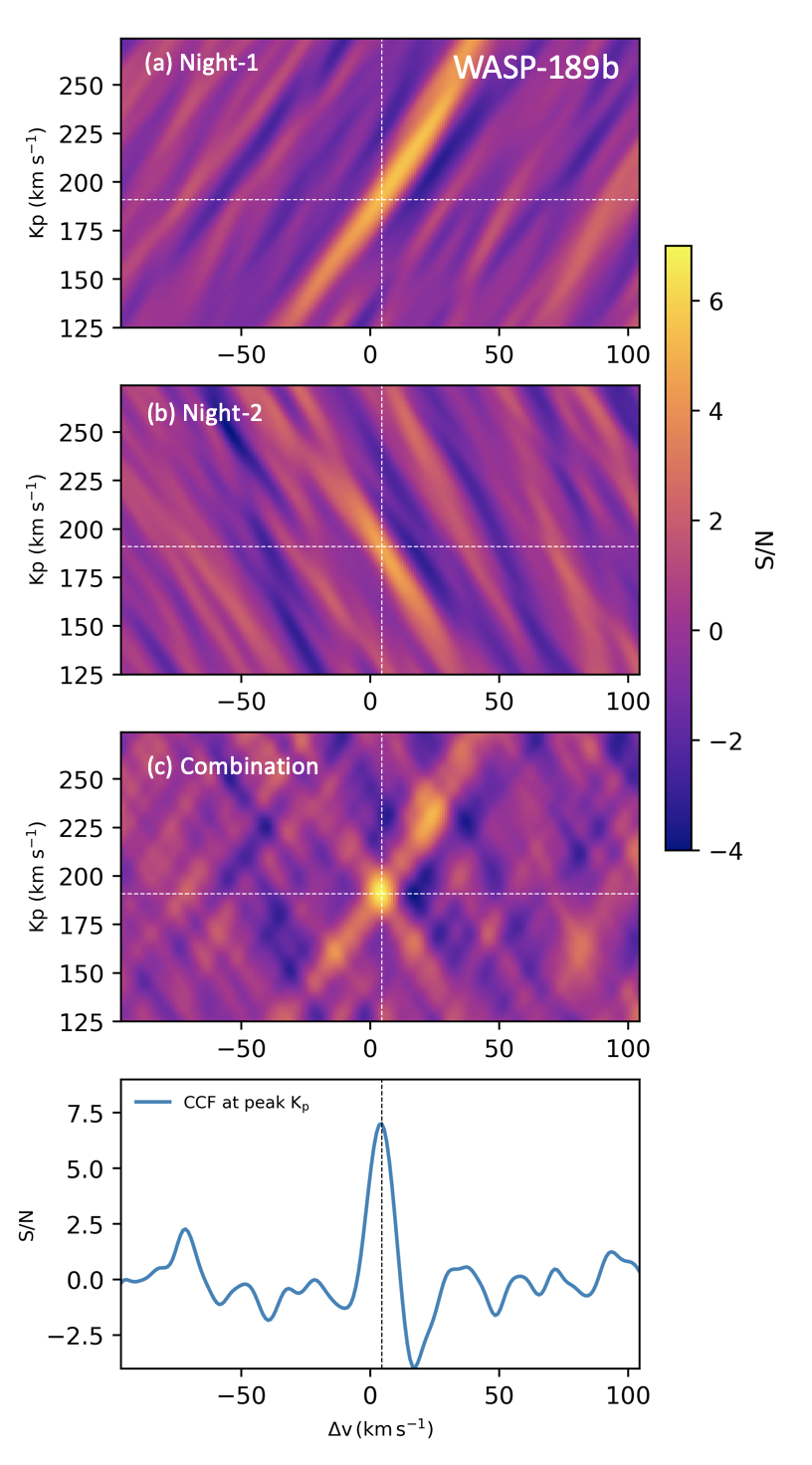}
      \caption{Same as Fig.~\ref{Kp-W33}, but for WASP-189b.}
         \label{Kp-W189}
   \end{figure}

To estimate the detection significance, we computed the $K_\mathrm{p}$ maps, where $K_\mathrm{p}$ is the semi-amplitude of the planetary orbital RV. The $K_\mathrm{p}$ maps were generated by combining all the out-of-eclipse CCFs in the planetary rest frame for different $K_\mathrm{p}$ values (ranging from 0 to 400 km\,s$^{-1}$ in steps of 1 km\,s$^{-1}$). We further divided the map by its noise, which was estimated as the standard deviation of the CCFs with $\left| \mathrm{\Delta} \varv \right|$ ranging from 100 to 200 km\,s$^{-1}$. In this way, we obtained the $K_\mathrm{p}$ maps in the unit of $\sigma$ (detection significance) (Figs.~\ref{Kp-W33} and \ref{Kp-W189}). We defined the horizontal axis of the map as $\mathrm{\Delta} \varv$, which means the RV deviation from the planetary rest frame (i.e., the remaining RV after the correction of the stellar systemic velocity and the planetary orbital motion RV).

For WASP-33b, we detected the CO signal with a S/N of 6.8 and the peak signal is located at $K_\mathrm{p}$ = $229_{-12}^{+13}$ km\,s$^{-1}$ and $\mathrm{\Delta} \varv$ = $2\pm10$ km\,s$^{-1}$. This $K_\mathrm{p}$ is consistent with the $K_\mathrm{p}$ inferred from the \ion{Fe}{i} emission lines ($225.0_{-3.5}^{+4.0}$ km\,s$^{-1}$) in \cite{Cont2021}.
For WASP-189b, the detected CO signal peaks at $K_\mathrm{p}$ = $191_{-5}^{+6}$ km\,s$^{-1}$ and $\mathrm{\Delta} \varv$ = $4.5_{-3.0}^{+2.2}$ km\,s$^{-1}$ with a S/N of 7.0. The detected $K_\mathrm{p}$ agrees well with the value from \ion{Fe}{i} emission lines ($193.54_{-0.57}^{+0.54}$ km\,s$^{-1}$) in \cite{Yan2020}.
The $K_\mathrm{p}$ results of both planets are also consistent with the expected $K_\mathrm{p}$ values calculated using the orbital parameters (cf. Tables \ref{paras_W33} and \ref{paras_W189}).

\subsection{Signature of atmospheric circulation}
\begin{figure}
   \centering
   \includegraphics[width=0.45\textwidth]{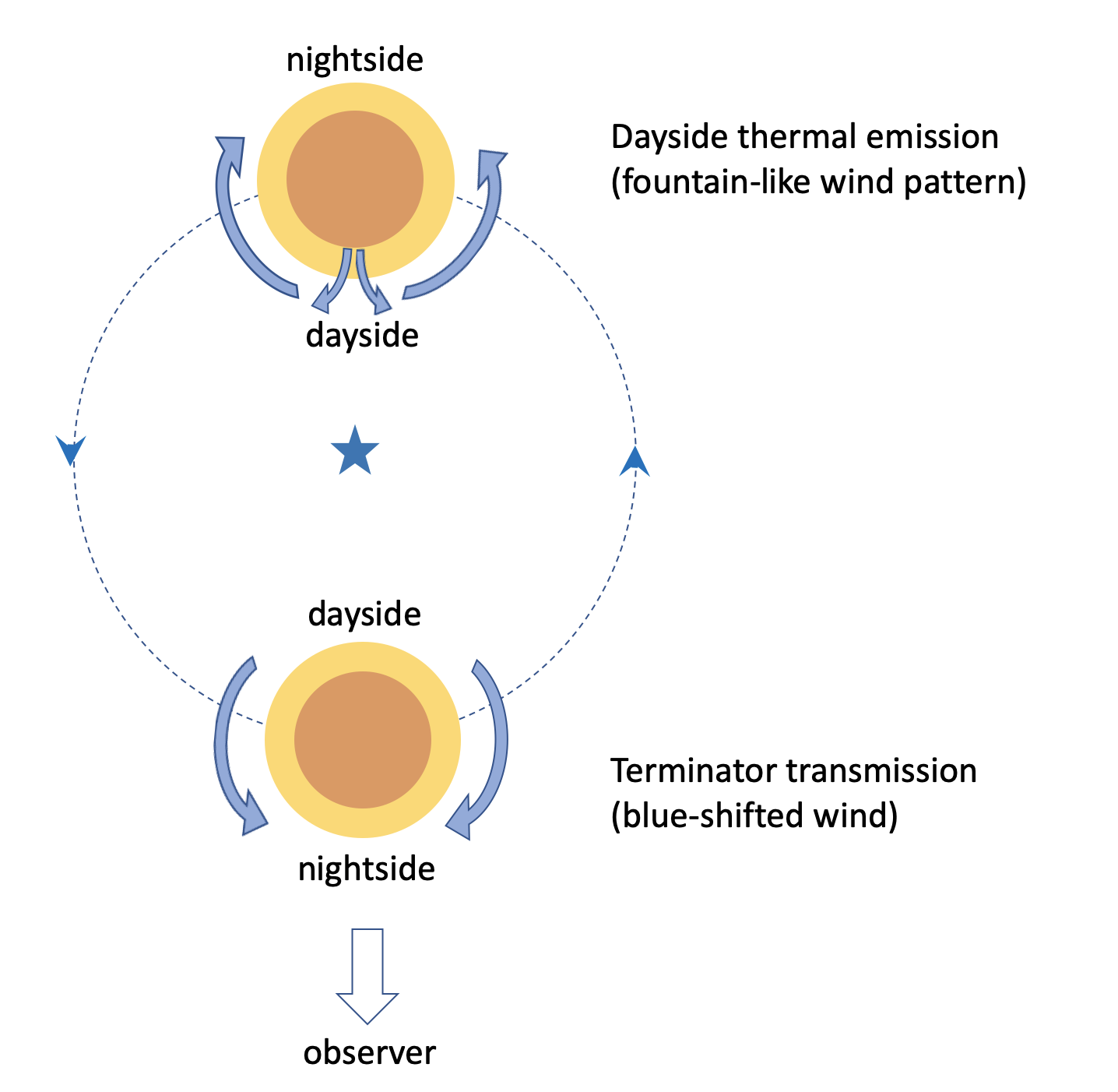}
      \caption{Schematic of the dayside to nightside wind for transmission and thermal emission observations. The dashed line denotes the planetary orbit. }
         \label{wind}
   \end{figure}

The detected CO signal of WASP-189b has a $\mathrm{\Delta} \varv$ of $4.5_{-3.0}^{+2.2}$ km\,s$^{-1}$, which was measured as the difference between the RV of the detected signal and the stellar systemic RV (i.e., the detected CO signal has an absolute RV of $-20_{-3.0}^{+2.2}$ km\,s$^{-1}$, while the stellar systemic velocity is $-24.452\pm0.012$ km\,s$^{-1}$).
The previously detected \ion{Fe}{i} emission lines also have a $\mathrm{\Delta} \varv$ of $4.3_{-0.26}^{+0.25}$ km\,s$^{-1}$ when using the same systemic velocity \citep{Yan2020}. Therefore, both the CO and \ion{Fe}{i} signals deviated from the systemic RV with a positive value, indicating that the dayside emission lines have a redshifted RV. Interestingly, recently detected transmission signals of WASP-189b \citep{Stangret2021, Prinoth2022} show a blueshifted RV of several km\,s$^{-1}$.
In particular, the transmission \ion{Fe}{i} lines have a $\mathrm{\Delta} \varv$ of --3.6 $\pm$ 0.4 km\,s$^{-1}$. 
Therefore, the $\mathrm{\Delta} \varv$ values of both the transmission and emission lines are significantly different from zero and the signs of the transmission and emission $\mathrm{\Delta} \varv$ values are opposite.
This can be explained consistently by assuming a dayside to nightside wind scenario \citep[e.g.,][]{Tan2019}.
For transmission observations, this scenario results in a blueshifted RV (Fig.~\ref{wind}). On the other hand, for thermal emission observations around secondary eclipses, the hotspot region on the dayside has a blueshifted RV because of the strong upward motion while the rest part of the dayside has a redshifted RV (cf. Fig.~\ref{wind} for such a fountain-like circulation pattern). Considering that the temperature of the hotspot is extremely high for WASP-189b, CO and \ion{Fe}{i} shall be largely disassociated or ionized at the center of the hotspot. Therefore, the center of the hotspot contributes less to the observed CO and \ion{Fe}{i} signals, while the rest of the dayside contributes more to the observed signals, leading to a redshifted RV of the emission lines.

For WASP-33b, such an RV signature of the dayside to nightside wind is not prominent. The detected $\mathrm{\Delta} \varv$ of the CO emission lines is $2\pm10$ km\,s$^{-1}$ and previously detected \ion{Fe}{i} emission lines have a $\mathrm{\Delta} \varv$ of $0.0\pm2.6$ km\,s$^{-1}$ \citep{Cont2021} and $0.2_{-1.8}^{+2.1}$ km\,s$^{-1}$ \citep{Nugroho2020W33}. The CO emission feature reported by \cite{vanSluijs-2022} also has a $\mathrm{\Delta} \varv$ of $0.15_{-0.65}^{+0.64}$ km\,s$^{-1}$. Therefore, there is no indication of a strong dayside to nightside wind in WASP-33b from thermal emission observations. For transmission signals, only \ion{Ca}{ii} and hydrogen Blamer lines have been detected in WASP-33b because the strong pulsation of the host star hinders the detection of atmospheric features. For the Blamer $\mathrm{H\alpha}$ line, different $\mathrm{\Delta} \varv$ values and line depths have been reported in \cite{Yan2021-W33} ($1.2\pm0.9$ km\,s$^{-1}$), \cite{Cauley2021-W33} ($-4.6\pm3.4$ km\,s$^{-1}$), and \cite{Borsa2021-W33} ($-8.2\pm1.4$ km\,s$^{-1}$), which is likely due to the strong contamination of stellar pulsations. The stellar systemic velocities used in these three works are slightly different, but the $\mathrm{\Delta} \varv$ values are still different when assuming the same systemic velocity.
Therefore, it is not conclusive on the existence of a dayside to nightside wind from the transmission observations.

According to the simulations of the atmospheric circulation in UHJs by \cite{Tan2019}, the atmospheric circulation patterns (e.g., the equatorial superrotation jet and the dayside to nightside wind) depend on multiple parameters.
In general, a longer rotation period, higher equilibrium temperature, and stronger frictional drag produce a stronger dayside to nightside wind with a weaker equatorial superrotation jet. The parameters of WASP-33b and WASP-189b are very similar (cf. Tables \ref{paras_W33} and \ref{paras_W189}); however, the rotation period (assumed to be tidally locked) of WASP-33b is half of the period of WASP-189b, which might be the reason that we observed a strong dayside to nightside wind in WASP-189b but not in WASP-33b. Such a theoretical explanation also suggests that WASP-33b may have a strong superrotation jet.

Dayside to nightside wind has also been observed in other UHJs such as WASP-76b \citep{Ehrenreich2020, Seidel2021} and HAT-P-70b \citep{Bello-Arufe2022} with transmission spectroscopy. Further studies of more UHJs with different parameters using both transmission and emission high-resolution spectroscopy will enable a systematic understanding of the atmospheric circulation in UHJs.

\section{Conclusions}
We observed the dayside thermal emission spectra of WASP-33b and WASP-189b with the high-resolution near-infrared spectrograph GIANO-B. Using the cross-correlation technique, we detected the CO emission lines at S/Ns of 6.8 and 7.0 for WASP-33b and WASP-189b, respectively.
The detected CO signals have a positive correlation with the spectral template that contains emission lines, meaning that the detected CO lines are in emission. This proves the existence of temperature inversion layers in their atmospheres, which agrees with the results from the previous detections of other emission lines (e.g., \ion{Fe}{i}) in the two UHJs. This is the first detection of CO lines in emission as previous high-resolution detections of CO lines are all in absorption.
The detected CO and \ion{Fe}{i} emission lines in WASP-189b have a redshifted RV of several km\,s$^{-1}$, which is likely due to the strong dayside to nightside wind in its atmosphere. Such a redshifted RV signature has not been observed in the emission lines of WASP-33b, hinting at different circulation patterns in the atmospheres of the two planets.

The detection of CO emission lines is a new addition to the discovered chemical species (i.e., \ion{Fe}{i}, \ion{Si}{i}, OH, and TiO) in the dayside atmosphere of UHJs with high-resolution spectroscopy. We note that CO emission features were previously inferred from the Spitzer photometry in several UHJs; therefore, the high-resolution detections strengthen the low-resolution evidences of CO signals. Further retrieval work combining both high- and low-resolution CO signals will enable a comprehensive understanding of the atmospheric properties such as temperature structures and C/O ratios.

Unlike other molecules, the thermal dissociation temperature of CO is relatively high. Thus, CO emission is likely to be a common spectral feature of UHJs. In addition, CO is an ideal species for tracing the planetary formation history via the measurement of the C/O ratio and the carbon isotope ratio. Therefore, CO lines are a promising spectral window into the understanding of the formation and migration of UHJs.

\begin{acknowledgements}
We thank the referee for the useful comments.
F.Y. acknowledges the support of the DFG Research Unit FOR2544 ``Blue Planets around Red Stars'' (RE 1664/21-1).
K.M. acknowledges the support of the Excellence Cluster ORIGINS which is funded by the Deutsche Forschungsgemeinschaft (DFG, German Research Foundation) under Germany's Excellence Strategy - EXC-2094 - 390783311.
P.M. acknowledges support from the European Research Council under the European Union's Horizon 2020 research and innovation program under grant agreement No. 832428-Origins. 
This work is based on observations made with the Italian Telescopio Nazionale Galileo (TNG) operated on the island of La Palma by the Fundaci\'on Galileo Galilei of the INAF (Istituto Nazionale di Astrofisica) at the Spanish Observatorio del Roque de los Muchachos of the Instituto de Astrofisica de Canarias. 
\end{acknowledgements}

\bibliographystyle{aa} 

\bibliography{UHJ-CO-refer}

\begin{thebibliography}{78}
\expandafter\ifx\csname natexlab\endcsname\relax\def\natexlab#1{#1}\fi

\bibitem[{{Anderson} {et~al.}(2018){Anderson}, {Temple}, {Nielsen}, {Burdanov},
  {Hellier}, {Bouchy}, {Brown}, {Collier Cameron}, {Gillon}, {Jehin}, {Maxted},
  {Pepe}, {Pollacco}, {Pozuelos}, {Queloz}, {S{\'e}gransan}, {Smalley},
  {Triaud}, {Turner}, {Udry}, \& {West}}]{Anderson2018}
{Anderson}, D.~R., {Temple}, L.~Y., {Nielsen}, L.~D., {et~al.} 2018, arXiv
  e-prints, arXiv:1809.04897

\bibitem[{{Bello-Arufe} {et~al.}(2022){Bello-Arufe}, {Cabot}, {Mendon{\c{c}}a},
  {Buchhave}, \& {Rathcke}}]{Bello-Arufe2022}
{Bello-Arufe}, A., {Cabot}, S. H.~C., {Mendon{\c{c}}a}, J.~M., {Buchhave},
  L.~A., \& {Rathcke}, A.~D. 2022, \aj, 163, 96

\bibitem[{{Ben-Yami} {et~al.}(2020){Ben-Yami}, {Madhusudhan}, {Cabot},
  {Constantinou}, {Piette}, {Gandhi}, \& {Welbanks}}]{Ben-Yami2020}
{Ben-Yami}, M., {Madhusudhan}, N., {Cabot}, S. H.~C., {et~al.} 2020, \apjl,
  897, L5

\bibitem[{{Birkby} {et~al.}(2013){Birkby}, {de Kok}, {Brogi}, {de Mooij},
  {Schwarz}, {Albrecht}, \& {Snellen}}]{Birkby2013}
{Birkby}, J.~L., {de Kok}, R.~J., {Brogi}, M., {et~al.} 2013, \mnras, 436, L35

\bibitem[{{Borsa} {et~al.}(2021{\natexlab{a}}){Borsa}, {Allart},
  {Casasayas-Barris}, {Tabernero}, {Zapatero Osorio}, {Cristiani}, {Pepe},
  {Rebolo}, {Santos}, {Adibekyan}, {Bourrier}, {Demangeon}, {Ehrenreich},
  {Pall{\'e}}, {Sousa}, {Lillo-Box}, {Lovis}, {Micela}, {Oshagh}, {Poretti},
  {Sozzetti}, {Allende Prieto}, {Alibert}, {Amate}, {Benz}, {Bouchy}, {Cabral},
  {Dekker}, {D'Odorico}, {Di Marcantonio}, {Figueira}, {Genova Santos},
  {Gonz{\'a}lez Hern{\'a}ndez}, {Lo Curto}, {Manescau}, {Martins},
  {M{\'e}gevand}, {Mehner}, {Molaro}, {Nunes}, {Riva}, {Su{\'a}rez
  Mascare{\~n}o}, {Udry}, \& {Zerbi}}]{Borsa2021-W121}
{Borsa}, F., {Allart}, R., {Casasayas-Barris}, N., {et~al.} 2021{\natexlab{a}},
  \aap, 645, A24

\bibitem[{{Borsa} {et~al.}(2021{\natexlab{b}}){Borsa}, {Fossati}, {Koskinen},
  {Young}, \& {Shulyak}}]{Borsa2021}
{Borsa}, F., {Fossati}, L., {Koskinen}, T., {Young}, M.~E., \& {Shulyak}, D.
  2021{\natexlab{b}}, Nature Astronomy

\bibitem[{{Borsa} {et~al.}(2021{\natexlab{c}}){Borsa}, {Lanza}, {Raspantini},
  {Rainer}, {Fossati}, {Brogi}, {Di Mauro}, {Gratton}, {Pino}, {Benatti},
  {Bignamini}, {Bonomo}, {Claudi}, {Esposito}, {Frustagli}, {Maggio},
  {Maldonado}, {Mancini}, {Micela}, {Nascimbeni}, {Poretti}, {Scandariato},
  {Sicilia}, {Sozzetti}, {Boschin}, {Cosentino}, {Covino}, {Desidera}, {Di
  Fabrizio}, {Fiorenzano}, {Harutyunyan}, {Knapic}, {Molinari}, {Pagano},
  {Pedani}, \& {Piotto}}]{Borsa2021-W33}
{Borsa}, F., {Lanza}, A.~F., {Raspantini}, I., {et~al.} 2021{\natexlab{c}},
  \aap, 653, A104

\bibitem[{{Bourrier} {et~al.}(2020){Bourrier}, {Ehrenreich}, {Lendl},
  {Cretignier}, {Allart}, {Dumusque}, {Cegla}, {Su{\'a}rez-Mascare{\~n}o},
  {Wyttenbach}, {Hoeijmakers}, {Melo}, {Kuntzer}, {Astudillo-Defru}, {Giles},
  {Heng}, {Kitzmann}, {Lavie}, {Lovis}, {Murgas}, {Nascimbeni}, {Pepe}, {Pino},
  {Segransan}, \& {Udry}}]{Bourrier2020}
{Bourrier}, V., {Ehrenreich}, D., {Lendl}, M., {et~al.} 2020, \aap, 635, A205

\bibitem[{{Brogi} {et~al.}(2018){Brogi}, {Giacobbe}, {Guilluy}, {de Kok},
  {Sozzetti}, {Mancini}, \& {Bonomo}}]{Brogi2018}
{Brogi}, M., {Giacobbe}, P., {Guilluy}, G., {et~al.} 2018, \aap, 615, A16

\bibitem[{{Brogi} {et~al.}(2012){Brogi}, {Snellen}, {de Kok}, {Albrecht},
  {Birkby}, \& {de Mooij}}]{Brogi2012}
{Brogi}, M., {Snellen}, I. A.~G., {de Kok}, R.~J., {et~al.} 2012, \nat, 486,
  502

\bibitem[{{Cabot} {et~al.}(2021){Cabot}, {Bello-Arufe}, {Mendon{\c{c}}a},
  {Tronsgaard}, {Wong}, {Zhou}, {Buchhave}, {Fischer}, {Stassun}, {Antoci},
  {Baker}, {Belinski}, {Benneke}, {Bouma}, {Christiansen}, {Collins},
  {Goliguzova}, {Hagey}, {Jenkins}, {Jensen}, {Kidwell}, {Laloum}, {Massey},
  {McLeod}, {Latham}, {Morgan}, {Ricker}, {Safonov}, {Schlieder}, {Seager},
  {Shporer}, {Smith}, {Srdoc}, {Strakhov}, {Torres}, {Twicken}, {Vanderspek},
  {Vezie}, \& {Winn}}]{Cabot2021}
{Cabot}, S. H.~C., {Bello-Arufe}, A., {Mendon{\c{c}}a}, J.~M., {et~al.} 2021,
  \aj, 162, 218

\bibitem[{{Cabot} {et~al.}(2020){Cabot}, {Madhusudhan}, {Welbanks}, {Piette},
  \& {Gandhi}}]{Cabot2020}
{Cabot}, S. H.~C., {Madhusudhan}, N., {Welbanks}, L., {Piette}, A., \&
  {Gandhi}, S. 2020, \mnras, 494, 363

\bibitem[{{Casasayas-Barris} {et~al.}(2021){Casasayas-Barris}, {Orell-Miquel},
  {Stangret}, {Nortmann}, {Yan}, {Oshagh}, {Palle}, {Sanz-Forcada},
  {L{\'o}pez-Puertas}, {Nagel}, {Luque}, {Morello}, {Snellen}, {Zechmeister},
  {Quirrenbach}, {Caballero}, {Ribas}, {Reiners}, {Amado}, {Bergond}, {Czesla},
  {Henning}, {Khalafinejad}, {Molaverdikhani}, {Montes}, {Perger},
  {S{\'a}nchez-L{\'o}pez}, \& {Sedaghati}}]{Casasayas-Barris2021}
{Casasayas-Barris}, N., {Orell-Miquel}, J., {Stangret}, M., {et~al.} 2021,
  \aap, 654, A163

\bibitem[{{Casasayas-Barris} {et~al.}(2018){Casasayas-Barris}, {Pall{\'e}},
  {Yan}, {Chen}, {Albrecht}, {Nortmann}, {Van Eylen}, {Snellen}, {Talens},
  {Gonz{\'a}lez Hern{\'a}ndez}, {Rebolo}, \& {Otten}}]{Casasayas-Barris2018}
{Casasayas-Barris}, N., {Pall{\'e}}, E., {Yan}, F., {et~al.} 2018, \aap, 616,
  A151

\bibitem[{{Casasayas-Barris} {et~al.}(2019){Casasayas-Barris}, {Pall{\'e}},
  {Yan}, {Chen}, {Kohl}, {Stangret}, {Parviainen}, {Helling}, {Watanabe},
  {Czesla}, {Fukui}, {Monta{\~n}{\'e}s-Rodr{\'\i}guez}, {Nagel}, {Narita},
  {Nortmann}, {Nowak}, {Schmitt}, \& {Zapatero Osorio}}]{Casasayas-Barris2019}
{Casasayas-Barris}, N., {Pall{\'e}}, E., {Yan}, F., {et~al.} 2019, \aap, 628,
  A9

\bibitem[{{Cauley} {et~al.}(2019){Cauley}, {Shkolnik}, {Ilyin}, {Strassmeier},
  {Redfield}, \& {Jensen}}]{Cauley2019}
{Cauley}, P.~W., {Shkolnik}, E.~L., {Ilyin}, I., {et~al.} 2019, \aj, 157, 69

\bibitem[{{Cauley} {et~al.}(2021){Cauley}, {Wang}, {Shkolnik}, {Ilyin},
  {Strassmeier}, {Redfield}, \& {Jensen}}]{Cauley2021-W33}
{Cauley}, P.~W., {Wang}, J., {Shkolnik}, E.~L., {et~al.} 2021, \aj, 161, 152

\bibitem[{{Claudi} {et~al.}(2017){Claudi}, {Benatti}, {Carleo}, {Ghedina},
  {Guerra}, {Micela}, {Molinari}, {Oliva}, {Rainer}, {Tozzi}, {Baffa},
  {Baruffolo}, {Buchschacher}, {Cecconi}, {Cosentino}, {Fantinel}, {Fini},
  {Ghinassi}, {Giani}, {Gonzalez}, {Gonzalez}, {Gratton}, {Harutyunyan},
  {Hernandez}, {Lodi}, {Malavolta}, {Maldonado}, {Origlia}, {Sanna}, {Sanjuan},
  {Scuderi}, {Seemann}, {Sozzetti}, {Perez Ventura}, {Hernandez Diaz}, {Galli},
  {Gonzalez}, {Riverol}, \& {Riverol}}]{Claudi2017}
{Claudi}, R., {Benatti}, S., {Carleo}, I., {et~al.} 2017, European Physical
  Journal Plus, 132, 364

\bibitem[{{Claudi} {et~al.}(2016){Claudi}, {Benatti}, {Carleo}, {Ghedina},
  {Molinari}, {Oliva}, {Tozzi}, {Baruffolo}, {Cecconi}, {Cosentino},
  {Fantinel}, {Fini}, {Ghinassi}, {Gonzalez}, {Gratton}, {Guerra},
  {Harutyunyan}, {Hernandez}, {Iuzzolino}, {Lodi}, {Malavolta}, {Maldonado},
  {Micela}, {Sanna}, {Sanjuan}, {Scuderi}, {Sozzetti}, {P{\'e}rez Ventura},
  {Diaz Marcos}, {Galli}, {Gonzalez}, {Riverol}, \& {Riverol}}]{Claudi2016}
{Claudi}, R., {Benatti}, S., {Carleo}, I., {et~al.} 2016, in Society of
  Photo-Optical Instrumentation Engineers (SPIE) Conference Series, Vol. 9908,
  Ground-based and Airborne Instrumentation for Astronomy VI, ed. C.~J.
  {Evans}, L.~{Simard}, \& H.~{Takami}, 99081A

\bibitem[{{Cont} {et~al.}(2021){Cont}, {Yan}, {Reiners}, {Casasayas-Barris},
  {Molli{\`e}re}, {Pall{\'e}}, {Henning}, {Nortmann}, {Stangret}, {Czesla},
  {L{\'o}pez-Puertas}, {S{\'a}nchez-L{\'o}pez}, {Rodler}, {Ribas},
  {Quirrenbach}, {Caballero}, {Amado}, {Carone}, {Khaimova}, {Kreidberg},
  {Molaverdikhani}, {Montes}, {Morello}, {Nagel}, {Oshagh}, \&
  {Zechmeister}}]{Cont2021}
{Cont}, D., {Yan}, F., {Reiners}, A., {et~al.} 2021, \aap, 651, A33

\bibitem[{{Cont} {et~al.}(2022){Cont}, {Yan}, {Reiners}, {Nortmann},
  {Molaverdikhani}, {Pall{\'e}}, {Stangret}, {Henning}, {Ribas}, {Quirrenbach},
  {Caballero}, {Zapatero Osorio}, {Amado}, {Aceituno}, {Casasayas-Barris},
  {Czesla}, {Kaminski}, {L{\'o}pez-Puertas}, {Montes}, {Morales}, {Morello},
  {Nagel}, {S{\'a}nchez-L{\'o}pez}, {Sedaghati}, \& {Zechmeister}}]{Cont2022}
{Cont}, D., {Yan}, F., {Reiners}, A., {et~al.} 2022, \aap, 657, L2

\bibitem[{{Deibert} {et~al.}(2021){Deibert}, {de Mooij}, {Jayawardhana},
  {Turner}, {Ridden-Harper}, {Fossati}, {Hood}, {Fortney}, {Flagg},
  {MacDonald}, {Allart}, \& {Sing}}]{Deibert2021}
{Deibert}, E.~K., {de Mooij}, E. J.~W., {Jayawardhana}, R., {et~al.} 2021,
  \apjl, 919, L15

\bibitem[{{Deline} {et~al.}(2022){Deline}, {Hooton}, {Lendl}, {Morris},
  {Salmon}, {Olofsson}, {Broeg}, {Ehrenreich}, {Beck}, {Brandeker}, {Hoyer},
  {Sulis}, {Van Grootel}, {Bourrier}, {Demangeon}, {Demory}, {Heng},
  {Parviainen}, {Serrano}, {Singh}, {Bonfanti}, {Fossati}, {Kitzmann}, {Sousa},
  {Wilson}, {Alibert}, {Alonso}, {Anglada}, {B{\'a}rczy}, {Barrado Navascues},
  {Barros}, {Baumjohann}, {Beck}, {Bekkelien}, {Benz}, {Billot}, {Bonfils},
  {Cabrera}, {Charnoz}, {Collier Cameron}, {Corral van Damme}, {Csizmadia},
  {Davies}, {Deleuil}, {Delrez}, {de Roche}, {Erikson}, {Fortier}, {Fridlund},
  {Futyan}, {Gandolfi}, {Gillon}, {G{\"u}del}, {Gutermann}, {Hasiba}, {Isaak},
  {Kiss}, {Laskar}, {Lecavelier des Etangs}, {Lovis}, {Magrin}, {Maxted},
  {Munari}, {Nascimbeni}, {Ottensamer}, {Pagano}, {Pall{\'e}}, {Peter},
  {Piotto}, {Pollacco}, {Queloz}, {Ragazzoni}, {Rando}, {Rauer}, {Ribas},
  {Santos}, {Scandariato}, {S{\'e}gransan}, {Simon}, {Smith}, {Steller},
  {Szab{\'o}}, {Thomas}, {Udry}, {Walter}, \& {Walton}}]{Deline2022}
{Deline}, A., {Hooton}, M.~J., {Lendl}, M., {et~al.} 2022, \aap, 659, A74

\bibitem[{{Ehrenreich} {et~al.}(2020){Ehrenreich}, {Lovis}, {Allart}, {Zapatero
  Osorio}, {Pepe}, {Cristiani}, {Rebolo}, {Santos}, {Borsa}, {Demangeon},
  {Dumusque}, {Gonz{\'a}lez Hern{\'a}ndez}, {Casasayas-Barris},
  {S{\'e}gransan}, {Sousa}, {Abreu}, {Adibekyan}, {Affolter}, {Allende Prieto},
  {Alibert}, {Aliverti}, {Alves}, {Amate}, {Avila}, {Baldini}, {Bandy}, {Benz},
  {Bianco}, {Bolmont}, {Bouchy}, {Bourrier}, {Broeg}, {Cabral}, {Calderone},
  {Pall{\'e}}, {Cegla}, {Cirami}, {Coelho}, {Conconi}, {Coretti}, {Cumani},
  {Cupani}, {Dekker}, {Delabre}, {Deiries}, {D'Odorico}, {Di Marcantonio},
  {Figueira}, {Fragoso}, {Genolet}, {Genoni}, {G{\'e}nova Santos}, {Hara},
  {Hughes}, {Iwert}, {Kerber}, {Knudstrup}, {Landoni}, {Lavie}, {Lizon},
  {Lendl}, {Lo Curto}, {Maire}, {Manescau}, {Martins}, {M{\'e}gevand},
  {Mehner}, {Micela}, {Modigliani}, {Molaro}, {Monteiro}, {Monteiro},
  {Moschetti}, {M{\"u}ller}, {Nunes}, {Oggioni}, {Oliveira}, {Pariani},
  {Pasquini}, {Poretti}, {Rasilla}, {Redaelli}, {Riva}, {Santana Tschudi},
  {Santin}, {Santos}, {Segovia Milla}, {Seidel}, {Sosnowska}, {Sozzetti},
  {Span{\`o}}, {Su{\'a}rez Mascare{\~n}o}, {Tabernero}, {Tenegi}, {Udry},
  {Zanutta}, \& {Zerbi}}]{Ehrenreich2020}
{Ehrenreich}, D., {Lovis}, C., {Allart}, R., {et~al.} 2020, \nat, 580, 597

\bibitem[{{Evans} {et~al.}(2017){Evans}, {Sing}, {Kataria}, {Goyal}, {Nikolov},
  {Wakeford}, {Deming}, {Marley}, {Amundsen}, {Ballester}, {Barstow},
  {Ben-Jaffel}, {Bourrier}, {Buchhave}, {Cohen}, {Ehrenreich}, {Garc{\'\i}a
  Mu{\~n}oz}, {Henry}, {Knutson}, {Lavvas}, {Lecavelier Des Etangs}, {Lewis},
  {L{\'o}pez-Morales}, {Mandell}, {Sanz-Forcada}, {Tremblin}, \&
  {Lupu}}]{Evans2017}
{Evans}, T.~M., {Sing}, D.~K., {Kataria}, T., {et~al.} 2017, \nat, 548, 58

\bibitem[{{Fossati} {et~al.}(2021){Fossati}, {Young}, {Shulyak}, {Koskinen},
  {Huang}, {Cubillos}, {France}, \& {Sreejith}}]{Fossati2021-NLTE}
{Fossati}, L., {Young}, M.~E., {Shulyak}, D., {et~al.} 2021, \aap, 653, A52

\bibitem[{{Fu} {et~al.}(2022){Fu}, {Sing}, {Lothringer}, {Deming}, {Ih},
  {Kempton}, {Malik}, {Komacek}, {Mansfield}, \& {Bean}}]{Fu2022}
{Fu}, G., {Sing}, D.~K., {Lothringer}, J.~D., {et~al.} 2022, \apjl, 925, L3

\bibitem[{{Giacobbe} {et~al.}(2021){Giacobbe}, {Brogi}, {Gandhi}, {Cubillos},
  {Bonomo}, {Sozzetti}, {Fossati}, {Guilluy}, {Carleo}, {Rainer},
  {Harutyunyan}, {Borsa}, {Pino}, {Nascimbeni}, {Benatti}, {Biazzo},
  {Bignamini}, {Chubb}, {Claudi}, {Cosentino}, {Covino}, {Damasso}, {Desidera},
  {Fiorenzano}, {Ghedina}, {Lanza}, {Leto}, {Maggio}, {Malavolta}, {Maldonado},
  {Micela}, {Molinari}, {Pagano}, {Pedani}, {Piotto}, {Poretti}, {Scandariato},
  {Yurchenko}, {Fantinel}, {Galli}, {Lodi}, {Sanna}, \& {Tozzi}}]{Giacobbe2021}
{Giacobbe}, P., {Brogi}, M., {Gandhi}, S., {et~al.} 2021, \nat, 592, 205

\bibitem[{{Gibson} {et~al.}(2020){Gibson}, {Merritt}, {Nugroho}, {Cubillos},
  {de Mooij}, {Mikal-Evans}, {Fossati}, {Lothringer}, {Nikolov}, {Sing},
  {Spake}, {Watson}, \& {Wilson}}]{Gibson2020}
{Gibson}, N.~P., {Merritt}, S., {Nugroho}, S.~K., {et~al.} 2020, \mnras, 493,
  2215

\bibitem[{{Guilluy} {et~al.}(2019){Guilluy}, {Sozzetti}, {Brogi}, {Bonomo},
  {Giacobbe}, {Claudi}, \& {Benatti}}]{Guilluy2019}
{Guilluy}, G., {Sozzetti}, A., {Brogi}, M., {et~al.} 2019, \aap, 625, A107

\bibitem[{{Helling} {et~al.}(2019){Helling}, {Gourbin}, {Woitke}, \&
  {Parmentier}}]{Helling2019}
{Helling}, C., {Gourbin}, P., {Woitke}, P., \& {Parmentier}, V. 2019, \aap,
  626, A133

\bibitem[{{Hoeijmakers} {et~al.}(2020{\natexlab{a}}){Hoeijmakers}, {Cabot},
  {Zhao}, {Buchhave}, {Tronsgaard}, {Davis}, {Kitzmann}, {Grimm}, {Cegla},
  {Bourrier}, {Ehrenreich}, {Heng}, {Lovis}, \& {Fischer}}]{Hoeijmakers2020}
{Hoeijmakers}, H.~J., {Cabot}, S. H.~C., {Zhao}, L., {et~al.}
  2020{\natexlab{a}}, \aap, 641, A120

\bibitem[{{Hoeijmakers} {et~al.}(2018){Hoeijmakers}, {Ehrenreich}, {Heng},
  {Kitzmann}, {Grimm}, {Allart}, {Deitrick}, {Wyttenbach}, {Oreshenko}, {Pino},
  {Rimmer}, {Molinari}, \& {Di Fabrizio}}]{Hoeijmakers2018}
{Hoeijmakers}, H.~J., {Ehrenreich}, D., {Heng}, K., {et~al.} 2018, \nat, 560,
  453

\bibitem[{{Hoeijmakers} {et~al.}(2019){Hoeijmakers}, {Ehrenreich}, {Kitzmann},
  {Allart}, {Grimm}, {Seidel}, {Wyttenbach}, {Pino}, {Nielsen}, {Fisher},
  {Rimmer}, {Bourrier}, {Cegla}, {Lavie}, {Lovis}, {Patzer}, {Stock}, {Pepe},
  \& {Heng}}]{Hoeijmakers2019}
{Hoeijmakers}, H.~J., {Ehrenreich}, D., {Kitzmann}, D., {et~al.} 2019, \aap,
  627, A165

\bibitem[{{Hoeijmakers} {et~al.}(2020{\natexlab{b}}){Hoeijmakers}, {Seidel},
  {Pino}, {Kitzmann}, {Sindel}, {Ehrenreich}, {Oza}, {Bourrier}, {Allart},
  {Gebek}, {Lovis}, {Yurchenko}, {Astudillo-Defru}, {Bayliss}, {Cegla},
  {Lavie}, {Lendl}, {Melo}, {Murgas}, {Nascimbeni}, {Pepe}, {S{\'e}gransan},
  {Udry}, {Wyttenbach}, \& {Heng}}]{Hoeijmakers2020-W121}
{Hoeijmakers}, H.~J., {Seidel}, J.~V., {Pino}, L., {et~al.} 2020{\natexlab{b}},
  \aap, 641, A123

\bibitem[{{Johnson} {et~al.}(2015){Johnson}, {Cochran}, {Collier Cameron}, \&
  {Bayliss}}]{Johnson2015}
{Johnson}, M.~C., {Cochran}, W.~D., {Collier Cameron}, A., \& {Bayliss}, D.
  2015, \apj, 810, L23

\bibitem[{{Kasper} {et~al.}(2021){Kasper}, {Bean}, {Line}, {Seifahrt},
  {St{\"u}rmer}, {Pino}, {D{\'e}sert}, \& {Brogi}}]{Kasper2021}
{Kasper}, D., {Bean}, J.~L., {Line}, M.~R., {et~al.} 2021, \apjl, 921, L18

\bibitem[{{Kesseli} {et~al.}(2022){Kesseli}, {Snellen}, {Casasayas-Barris},
  {Molli{\`e}re}, \& {S{\'a}nchez-L{\'o}pez}}]{Kesseli2022}
{Kesseli}, A.~Y., {Snellen}, I.~A.~G., {Casasayas-Barris}, N., {Molli{\`e}re},
  P., \& {S{\'a}nchez-L{\'o}pez}, A. 2022, \aj, 163, 107

\bibitem[{{Kitzmann} {et~al.}(2018){Kitzmann}, {Heng}, {Rimmer}, {Hoeijmakers},
  {Tsai}, {Malik}, {Lendl}, {Deitrick}, \& {Demory}}]{Kitzmann2018}
{Kitzmann}, D., {Heng}, K., {Rimmer}, P.~B., {et~al.} 2018, \apj, 863, 183

\bibitem[{{Konopacky} {et~al.}(2013){Konopacky}, {Barman}, {Macintosh}, \&
  {Marois}}]{Konopacky2013}
{Konopacky}, Q.~M., {Barman}, T.~S., {Macintosh}, B.~A., \& {Marois}, C. 2013,
  Science, 339, 1398

\bibitem[{{Kreidberg} {et~al.}(2018){Kreidberg}, {Line}, {Parmentier},
  {Stevenson}, {Louden}, {Bonnefoy}, {Faherty}, {Henry}, {Williamson},
  {Stassun}, {Beatty}, {Bean}, {Fortney}, {Showman}, {D{\'e}sert}, \&
  {Arcangeli}}]{Kreidberg2018}
{Kreidberg}, L., {Line}, M.~R., {Parmentier}, V., {et~al.} 2018, \aj, 156, 17

\bibitem[{{Landman} {et~al.}(2021){Landman}, {S{\'a}nchez-L{\'o}pez},
  {Molli{\`e}re}, {Kesseli}, {Louca}, \& {Snellen}}]{Landman2021}
{Landman}, R., {S{\'a}nchez-L{\'o}pez}, A., {Molli{\`e}re}, P., {et~al.} 2021,
  \aap, 656, A119

\bibitem[{{Lehmann} {et~al.}(2015){Lehmann}, {Guenther}, {Sebastian},
  {D{\"o}llinger}, {Hartmann}, \& {Mkrtichian}}]{Lehmann2015}
{Lehmann}, H., {Guenther}, E., {Sebastian}, D., {et~al.} 2015, \aap, 578, L4

\bibitem[{{Li} {et~al.}(2015){Li}, {Gordon}, {Rothman}, {Tan}, {Hu}, {Kassi},
  {Campargue}, \& {Medvedev}}]{Li2015}
{Li}, G., {Gordon}, I.~E., {Rothman}, L.~S., {et~al.} 2015, \apjs, 216, 15

\bibitem[{{Line} {et~al.}(2021){Line}, {Brogi}, {Bean}, {Gandhi}, {Zalesky},
  {Parmentier}, {Smith}, {Mace}, {Mansfield}, {Kempton}, {Fortney}, {Shkolnik},
  {Patience}, {Rauscher}, {D{\'e}sert}, \& {Wardenier}}]{Line2021}
{Line}, M.~R., {Brogi}, M., {Bean}, J.~L., {et~al.} 2021, \nat, 598, 580

\bibitem[{{Lothringer} {et~al.}(2018){Lothringer}, {Barman}, \&
  {Koskinen}}]{Lothringer2018}
{Lothringer}, J.~D., {Barman}, T., \& {Koskinen}, T. 2018, \apj, 866, 27

\bibitem[{{Maciejewski} {et~al.}(2018){Maciejewski}, {Fern{\'a}ndez},
  {Aceituno}, {Mart{\'\i}n-Ruiz}, {Ohlert}, {Dimitrov}, {Szyszka}, {von Essen},
  {Mugrauer}, {Bischoff}, {Michel}, {Mallonn}, {Stangret}, \&
  {Mo{\'z}dzierski}}]{Maciejewski2018}
{Maciejewski}, G., {Fern{\'a}ndez}, M., {Aceituno}, F., {et~al.} 2018, \actaa,
  68, 371

\bibitem[{{Molli{\`e}re} {et~al.}(2015){Molli{\`e}re}, {van Boekel},
  {Dullemond}, {Henning}, \& {Mordasini}}]{Molliere2015}
{Molli{\`e}re}, P., {van Boekel}, R., {Dullemond}, C., {Henning}, T., \&
  {Mordasini}, C. 2015, \apj, 813, 47

\bibitem[{{Molli{\`e}re} {et~al.}(2019){Molli{\`e}re}, {Wardenier}, {van
  Boekel}, {Henning}, {Molaverdikhani}, \& {Snellen}}]{Molliere2019}
{Molli{\`e}re}, P., {Wardenier}, J.~P., {van Boekel}, R., {et~al.} 2019, \aap,
  627, A67

\bibitem[{{Nugroho} {et~al.}(2020){Nugroho}, {Gibson}, {de Mooij}, {Herman},
  {Watson}, {Kawahara}, \& {Merritt}}]{Nugroho2020W33}
{Nugroho}, S.~K., {Gibson}, N.~P., {de Mooij}, E. J.~W., {et~al.} 2020, \apjl,
  898, L31

\bibitem[{{Nugroho} {et~al.}(2021){Nugroho}, {Kawahara}, {Gibson}, {de Mooij},
  {Hirano}, {Kotani}, {Kawashima}, {Masuda}, {Brogi}, {Birkby}, {Watson},
  {Tamura}, {Zwintz}, {Harakawa}, {Kudo}, {Kuzuhara}, {Hodapp}, {Ishizuka},
  {Jacobson}, {Konishi}, {Kurokawa}, {Nishikawa}, {Omiya}, {Serizawa}, {Ueda},
  \& {Vievard}}]{Nugroho2021}
{Nugroho}, S.~K., {Kawahara}, H., {Gibson}, N.~P., {et~al.} 2021, \apjl, 910,
  L9

\bibitem[{{Nugroho} {et~al.}(2017){Nugroho}, {Kawahara}, {Masuda}, {Hirano},
  {Kotani}, \& {Tajitsu}}]{Nugroho2017}
{Nugroho}, S.~K., {Kawahara}, H., {Masuda}, K., {et~al.} 2017, \aj, 154, 221

\bibitem[{{Pai Asnodkar} {et~al.}(2022){Pai Asnodkar}, {Wang}, {Eastman},
  {Cauley}, {Gaudi}, {Ilyin}, \& {Strassmeier}}]{PaiAsnodkar2022}
{Pai Asnodkar}, A., {Wang}, J., {Eastman}, J.~D., {et~al.} 2022, \aj, 163, 155

\bibitem[{{Pino} {et~al.}(2020){Pino}, {D{\'e}sert}, {Brogi}, {Malavolta},
  {Wyttenbach}, {Line}, {Hoeijmakers}, {Fossati}, {Bonomo}, {Nascimbeni},
  {Panwar}, {Affer}, {Benatti}, {Biazzo}, {Bignamini}, {Borsa}, {Carleo},
  {Claudi}, {Cosentino}, {Covino}, {Damasso}, {Desidera}, {Giacobbe},
  {Harutyunyan}, {Lanza}, {Leto}, {Maggio}, {Maldonado}, {Mancini}, {Micela},
  {Molinari}, {Pagano}, {Piotto}, {Poretti}, {Rainer}, {Scandariato},
  {Sozzetti}, {Allart}, {Borsato}, {Bruno}, {Fabrizio}, {Ehrenreich},
  {Fiorenzano}, {Frustagli}, {Lavie}, {Lovis}, {Magazz{\`u}}, {Nardiello},
  {Pedani}, \& {Smareglia}}]{Pino2020}
{Pino}, L., {D{\'e}sert}, J.-M., {Brogi}, M., {et~al.} 2020, \apjl, 894, L27

\bibitem[{{Prinoth} {et~al.}(2022){Prinoth}, {Hoeijmakers}, {Kitzmann},
  {Sandvik}, {Seidel}, {Lendl}, {Borsato}, {Thorsbro}, {Anderson}, {Barrado},
  {Kravchenko}, {Allart}, {Bourrier}, {Cegla}, {Ehrenreich}, {Fisher}, {Lovis},
  {Guzm{\'a}n-Mesa}, {Grimm}, {Hooton}, {Morris}, {Oreshenko}, {Pino}, \&
  {Heng}}]{Prinoth2022}
{Prinoth}, B., {Hoeijmakers}, H.~J., {Kitzmann}, D., {et~al.} 2022, Nature
  Astronomy [\eprint[arXiv]{2111.12732}]

\bibitem[{{Rainer} {et~al.}(2021){Rainer}, {Borsa}, {Pino}, {Frustagli},
  {Brogi}, {Biazzo}, {Bonomo}, {Carleo}, {Claudi}, {Gratton}, {Lanza},
  {Maggio}, {Maldonado}, {Mancini}, {Micela}, {Scandariato}, {Sozzetti},
  {Buchschacher}, {Cosentino}, {Covino}, {Ghedina}, {Gonzalez}, {Leto}, {Lodi},
  {Martinez Fiorenzano}, {Molinari}, {Molinaro}, {Nardiello}, {Oliva},
  {Pagano}, {Pedani}, {Piotto}, \& {Poretti}}]{Rainer2021}
{Rainer}, M., {Borsa}, F., {Pino}, L., {et~al.} 2021, \aap, 649, A29

\bibitem[{{Rainer} {et~al.}(2018){Rainer}, {Harutyunyan}, {Carleo}, {Oliva},
  {Benatti}, {Bignamini}, {Claudi}, {Gonzalez-Alvarez}, {Sanna}, {Ghedina},
  {Micela}, {Molinari}, {Tozzi}, {Baffa}, {Baruffolo}, {Buchschacher},
  {Cecconi}, {Cosentino}, {Falcini}, {Fantinel}, {Fini}, {Galli}, {Ghinassi},
  {Giani}, {Gonzalez}, {Gonzalez}, {Gratton}, {Guerra}, {Hernandez Diaz},
  {Hernandez}, {Iuzzolino}, {Lodi}, {Malavolta}, {Maldonado}, {Origlia}, {Perez
  Ventura}, {Puglisi}, {Riverol}, {Riverol}, {San Juan}, {Scuderi}, {Seeman},
  {Sozzetti}, \& {Sozzi}}]{Rainer2018}
{Rainer}, M., {Harutyunyan}, A., {Carleo}, I., {et~al.} 2018, in Society of
  Photo-Optical Instrumentation Engineers (SPIE) Conference Series, Vol. 10702,
  Ground-based and Airborne Instrumentation for Astronomy VII, ed. C.~J.
  {Evans}, L.~{Simard}, \& H.~{Takami}, 1070266

\bibitem[{{Seidel} {et~al.}(2021){Seidel}, {Ehrenreich}, {Allart},
  {Hoeijmakers}, {Lovis}, {Bourrier}, {Pino}, {Wyttenbach}, {Adibekyan},
  {Alibert}, {Borsa}, {Casasayas-Barris}, {Cristiani}, {Demangeon}, {Di
  Marcantonio}, {Figueira}, {Gonz{\'a}lez Hern{\'a}ndez}, {Lillo-Box},
  {Martins}, {Mehner}, {Molaro}, {Nunes}, {Palle}, {Pepe}, {Santos}, {Sousa},
  {Sozzetti}, {Tabernero}, \& {Zapatero Osorio}}]{Seidel2021}
{Seidel}, J.~V., {Ehrenreich}, D., {Allart}, R., {et~al.} 2021, \aap, 653, A73

\bibitem[{{Seidel} {et~al.}(2019){Seidel}, {Ehrenreich}, {Wyttenbach},
  {Allart}, {Lendl}, {Pino}, {Bourrier}, {Cegla}, {Lovis}, {Barrado},
  {Bayliss}, {Astudillo-Defru}, {Deline}, {Fisher}, {Heng}, {Joseph}, {Lavie},
  {Melo}, {Pepe}, {S{\'e}gransan}, \& {Udry}}]{Seidel2019}
{Seidel}, J.~V., {Ehrenreich}, D., {Wyttenbach}, A., {et~al.} 2019, \aap, 623,
  A166

\bibitem[{{Sheppard} {et~al.}(2017){Sheppard}, {Mandell}, {Tamburo}, {Gand hi},
  {Pinhas}, {Madhusudhan}, \& {Deming}}]{Sheppard2017}
{Sheppard}, K.~B., {Mandell}, A.~M., {Tamburo}, P., {et~al.} 2017, \apjl, 850,
  L32

\bibitem[{{Sing} {et~al.}(2019){Sing}, {Lavvas}, {Ballester}, {Lecavelier des
  Etangs}, {Marley}, {Nikolov}, {Ben-Jaffel}, {Bourrier}, {Buchhave}, {Deming},
  {Ehrenreich}, {Mikal-Evans}, {Kataria}, {Lewis}, {L{\'o}pez-Morales},
  {Garc{\'\i}a Mu{\~n}oz}, {Henry}, {Sanz-Forcada}, {Spake}, {Wakeford}, \&
  {PanCET Collaboration}}]{Sing2019}
{Sing}, D.~K., {Lavvas}, P., {Ballester}, G.~E., {et~al.} 2019, \aj, 158, 91

\bibitem[{{Smette} {et~al.}(2015){Smette}, {Sana}, {Noll}, {Horst}, {Kausch},
  {Kimeswenger}, {Barden}, {Szyszka}, {Jones}, {Gallenne}, {Vinther},
  {Ballester}, \& {Taylor}}]{Smette2015}
{Smette}, A., {Sana}, H., {Noll}, S., {et~al.} 2015, \aap, 576, A77

\bibitem[{{Snellen} {et~al.}(2010){Snellen}, {de Kok}, {de Mooij}, \&
  {Albrecht}}]{Snellen2010}
{Snellen}, I.~A.~G., {de Kok}, R.~J., {de Mooij}, E.~J.~W., \& {Albrecht}, S.
  2010, \nat, 465, 1049

\bibitem[{{Stangret} {et~al.}(2021){Stangret}, {Casasayas-Barris}, {Pall{\'e}},
  {Orell-Miquel}, {Morello}, {Luque}, {Nowak}, \& {Yan}}]{Stangret2021}
{Stangret}, M., {Casasayas-Barris}, N., {Pall{\'e}}, E., {et~al.} 2021, arXiv
  e-prints, arXiv:2111.13064

\bibitem[{{Stangret} {et~al.}(2020){Stangret}, {Casasayas-Barris}, {Pall{\'e}},
  {Yan}, {S{\'a}nchez-L{\'o}pez}, \& {L{\'o}pez-Puertas}}]{Stangret2020}
{Stangret}, M., {Casasayas-Barris}, N., {Pall{\'e}}, E., {et~al.} 2020, \aap,
  638, A26

\bibitem[{{Tabernero} {et~al.}(2021){Tabernero}, {Zapatero Osorio}, {Allart},
  {Borsa}, {Casasayas-Barris}, {Demangeon}, {Ehrenreich}, {Lillo-Box}, {Lovis},
  {Pall{\'e}}, {Sousa}, {Rebolo}, {Santos}, {Pepe}, {Cristiani}, {Adibekyan},
  {Allende Prieto}, {Alibert}, {Barros}, {Bouchy}, {Bourrier}, {D'Odorico},
  {Dumusque}, {Faria}, {Figueira}, {G{\'e}nova Santos}, {Gonz{\'a}lez
  Hern{\'a}ndez}, {Hojjatpanah}, {Lo Curto}, {Lavie}, {Martins}, {Martins},
  {Mehner}, {Micela}, {Molaro}, {Nunes}, {Poretti}, {Seidel}, {Sozzetti},
  {Su{\'a}rez Mascare{\~n}o}, {Udry}, {Aliverti}, {Affolter}, {Alves}, {Amate},
  {Avila}, {Bandy}, {Benz}, {Bianco}, {Broeg}, {Cabral}, {Conconi}, {Coelho},
  {Cumani}, {Deiries}, {Dekker}, {Delabre}, {Fragoso}, {Genoni}, {Genolet},
  {Hughes}, {Knudstrup}, {Kerber}, {Landoni}, {Lizon}, {Maire}, {Manescau}, {Di
  Marcantonio}, {M{\'e}gevand}, {Monteiro}, {Monteiro}, {Moschetti}, {Mueller},
  {Modigliani}, {Oggioni}, {Oliveira}, {Pariani}, {Pasquini}, {Rasilla},
  {Redaelli}, {Riva}, {Santana-Tschudi}, {Santin}, {Santos}, {Segovia},
  {Sosnowska}, {Span{\`o}}, {Tenegi}, {Iwert}, {Zanutta}, \&
  {Zerbi}}]{Tabernero2021}
{Tabernero}, H.~M., {Zapatero Osorio}, M.~R., {Allart}, R., {et~al.} 2021,
  \aap, 646, A158

\bibitem[{{Tamuz} {et~al.}(2005){Tamuz}, {Mazeh}, \& {Zucker}}]{Tamuz2005}
{Tamuz}, O., {Mazeh}, T., \& {Zucker}, S. 2005, \mnras, 356, 1466

\bibitem[{{Tan} \& {Komacek}(2019)}]{Tan2019}
{Tan}, X. \& {Komacek}, T.~D. 2019, \apj, 886, 26

\bibitem[{{Turner} {et~al.}(2020){Turner}, {de Mooij}, {Jayawardhana}, {Young},
  {Fossati}, {Koskinen}, {Lothringer}, {Karjalainen}, \&
  {Karjalainen}}]{Turner2020}
{Turner}, J.~D., {de Mooij}, E. J.~W., {Jayawardhana}, R., {et~al.} 2020,
  \apjl, 888, L13

\bibitem[{{van Sluijs} {et~al.}(2022){van Sluijs}, {Birkby}, {Lothringer},
  {Lee}, {Crossfield}, {Parmentier}, {Brogi}, {Kulesa}, {McCarthy}, {Powell},
  \& {Charbonneau}}]{vanSluijs-2022}
{van Sluijs}, L., {Birkby}, J.~L., {Lothringer}, J., {et~al.} 2022, arXiv
  e-prints, arXiv:2203.13234

\bibitem[{{{\v{Z}}{\'a}k} {et~al.}(2019){{\v{Z}}{\'a}k}, {Kab{\'a}th},
  {Boffin}, {Ivanov}, \& {Skarka}}]{Zak2019}
{{\v{Z}}{\'a}k}, J., {Kab{\'a}th}, P., {Boffin}, H. M.~J., {Ivanov}, V.~D., \&
  {Skarka}, M. 2019, \aj, 158, 120

\bibitem[{{Wyttenbach} {et~al.}(2020){Wyttenbach}, {Molli{\`e}re},
  {Ehrenreich}, {Cegla}, {Bourrier}, {Lovis}, {Pino}, {Allart}, {Seidel},
  {Hoeijmakers}, {Nielsen}, {Lavie}, {Pepe}, {Bonfils}, \&
  {Snellen}}]{Wyttenbach2020}
{Wyttenbach}, A., {Molli{\`e}re}, P., {Ehrenreich}, D., {et~al.} 2020, \aap,
  638, A87

\bibitem[{{Yan} {et~al.}(2019){Yan}, {Casasayas-Barris}, {Molaverdikhani},
  {Alonso-Floriano}, {Reiners}, {Pall{\'e}}, {Henning}, {Molli{\`e}re}, {Chen},
  {Nortmann}, {Snellen}, {Ribas}, {Quirrenbach}, {Caballero}, {Amado},
  {Azzaro}, {Bauer}, {Cort{\'e}s Contreras}, {Czesla}, {Khalafinejad}, {Lara},
  {L{\'o}pez-Puertas}, {Montes}, {Nagel}, {Oshagh}, {S{\'a}nchez-L{\'o}pez},
  {Stangret}, \& {Zechmeister}}]{Yan2019}
{Yan}, F., {Casasayas-Barris}, N., {Molaverdikhani}, K., {et~al.} 2019, \aap,
  632, A69

\bibitem[{{Yan} \& {Henning}(2018)}]{Yan2018}
{Yan}, F. \& {Henning}, T. 2018, Nature Astronomy, 2, 714

\bibitem[{{Yan} {et~al.}(2020){Yan}, {Pall{\'e}}, {Reiners}, {Molaverdikhani},
  {Casasayas-Barris}, {Nortmann}, {Chen}, {Molli{\`e}re}, \&
  {Stangret}}]{Yan2020}
{Yan}, F., {Pall{\'e}}, E., {Reiners}, A., {et~al.} 2020, \aap, 640, L5

\bibitem[{{Yan} {et~al.}(2022){Yan}, {Reiners}, {Pall{\'e}}, {Shulyak},
  {Stangret}, {Molaverdikhani}, {Nortmann}, {Molli{\`e}re}, {Henning},
  {Casasayas-Barris}, {Cont}, {Chen}, {Czesla}, {S{\'a}nchez-L{\'o}pez},
  {L{\'o}pez-Puertas}, {Ribas}, {Quirrenbach}, {Caballero}, {Amado},
  {Galad{\'\i}-Enr{\'\i}quez}, {Khalafinejad}, {Lara}, {Montes}, {Morello},
  {Nagel}, {Sedaghati}, {Zapatero Osorio}, \& {Zechmeister}}]{Yan2022}
{Yan}, F., {Reiners}, A., {Pall{\'e}}, E., {et~al.} 2022, \aap, 659, A7

\bibitem[{{Yan} {et~al.}(2021){Yan}, {Wyttenbach}, {Casasayas-Barris},
  {Reiners}, {Pall{\'e}}, {Henning}, {Molli{\`e}re}, {Czesla}, {Nortmann},
  {Molaverdikhani}, {Chen}, {Snellen}, {Zechmeister}, {Huang}, {Ribas},
  {Quirrenbach}, {Caballero}, {Amado}, {Cont}, {Khalafinejad}, {Khaimova},
  {L{\'o}pez-Puertas}, {Montes}, {Nagel}, {Oshagh}, {Pedraz}, \&
  {Stangret}}]{Yan2021-W33}
{Yan}, F., {Wyttenbach}, A., {Casasayas-Barris}, N., {et~al.} 2021, \aap, 645,
  A22

\bibitem[{{Yurchenko} {et~al.}(2018){Yurchenko}, {Al-Refaie}, \&
  {Tennyson}}]{Yurchenko2018}
{Yurchenko}, S.~N., {Al-Refaie}, A.~F., \& {Tennyson}, J. 2018, \aap, 614, A131

\end{thebibliography}

\begin{appendix}
\section{Additional tables and figures}

%
\begin{table}
\small 
\begin{threeparttable}                       
\caption{Parameters of the WASP-33b system.}             
\label{paras_W33}      
\centering   
\begin{tabular}{l c c  }        
\hline\hline    \noalign{\smallskip}             
        Parameter & Symbol [unit] & Value   \\
        \hline \noalign{\smallskip}                      
        \textit{Star} & ~  & ~ \\
        Effective temperature & $T_\mathrm{eff}$ [K]  &       7430 $\pm$ 100 \tnote{a}   \\
        Radius & $R_\star$ [$R_\odot$] &       1.509$_{-0.027}^{+0.016}$ \tnote{a}    \\ \noalign{\smallskip}
        Mass & $M_\star$     [$M_\odot$]&    1.561$_{-0.079}^{+0.045}$ \tnote{a}    \\ \noalign{\smallskip}
        Systemic velocity &$v_\mathrm{sys}$ [km\,s$^{-1}$]     &       --3.0 $\pm$ 0.4 \tnote{b}   \\                 
         ~ & ~  \\
        \textit{Planet} & ~ & ~  \\
        Radius & $R_\mathrm{p}$ [$R_\mathrm{J}$]       & 1.679$_{-0.030}^{+0.019}$ \tnote{a}    \\ \noalign{\smallskip}
        Mass & $M_\mathrm{p}$        [$M_\mathrm{J}$]        &       2.16 $\pm$ 0.20 \tnote{a}    \\
        Semi-major axis & $a$   [$R_\star$]      &      3.69 $\pm$ 0.05 \tnote{a}   \\
        Orbital period & $P$ [d]        &       1.219870897(79) \tnote{c}  \\
        Transit epoch (BJD) & $T_\mathrm {0}$ [d]       &       2454163.22449(16) \tnote{c}   \\
        RV semi-amplitude &$K_\mathrm{p}$ [km\,s$^{-1}$] & 231 $\pm$ 3 \tnote{a}    \\
        Equilibrium temperature &$T_\mathrm{eq}$ [K] &  2710 $\pm$ 50 \\
        Orbital inclination &$i$ [deg]        &       89.2 \tnote{d}    \\
\hline                                   
\end{tabular}
\tablefoot{
  \tablefoottext{a}{\cite{Lehmann2015}.}
  \tablefoottext{b}{\cite{Nugroho2017}.}
  \tablefoottext{c}{\cite{Maciejewski2018}.}
  \tablefoottext{d}{Predicted value using orbital precession parameters in \cite{Johnson2015}.}  
}
\end{threeparttable}      
\end{table}

%
\begin{table}
\small 
\begin{threeparttable}                       
\caption{Parameters of the WASP-189b system.}             
\label{paras_W189}      
\centering   
\begin{tabular}{l c c  }        
\hline\hline    \noalign{\smallskip}             
        Parameter & Symbol [unit] & Value   \\
        \hline \noalign{\smallskip}                      
        \textit{Star} & ~  & ~ \\
        Effective temperature & $T_\mathrm{eff}$ [K]  &       8000 $\pm$ 80 \tnote{a}   \\
        Radius & $R_\star$ [$R_\odot$] &       2.365 $\pm$ 0.025 \tnote{a}    \\ \noalign{\smallskip}
        Mass & $M_\star$     [$M_\odot$]&    2.031$_{-0.096}^{+0.098}$ \tnote{a}    \\ \noalign{\smallskip}
        Systemic velocity &$v_\mathrm{sys}$ [km\,s$^{-1}$]     &       --24.452 $\pm$ 0.012 \tnote{b}   \\                 
         ~ & ~  \\
        \textit{Planet} & ~ & ~  \\
        Radius & $R_\mathrm{p}$ [$R_\mathrm{J}$]       & 1.600$_{-0.016}^{+0.017}$ \tnote{a}    \\ \noalign{\smallskip}
        Mass & $M_\mathrm{p}$        [$M_\mathrm{J}$]        &       2.13 $\pm$ 0.28 \tnote{b}    \\
        Semi-major axis & $a$   [$R_\star$]      &      4.587$_{-0.034}^{+0.037}$ \tnote{a}   \\ \noalign{\smallskip}
        Orbital period & $P$ [d]        &       2.724035(23) \tnote{a}  \\
        Transit epoch (BJD) & $T_\mathrm {0}$ [d]       &       2459016.43487(6) \tnote{a}   \\
        RV semi-amplitude & $K_\mathrm{p}$ [km\,s$^{-1}$] & 192 $\pm$ 3 \tnote{a}    \\
        ~ & ~ & 197$_{-11}^{+10}$ \tnote{b}    \\
        Equilibrium temperature &$T_\mathrm{eq}$ [K] &  2641 $\pm$ 34  \tnote{b} \\
        Orbital inclination &$i$ [deg]        &       84.58$_{-0.22}^{+0.23}$ \tnote{a}    \\ \noalign{\smallskip}
\hline                                   
\end{tabular}
\tablefoot{
  \tablefoottext{a}{\cite{Deline2022}.}
  \tablefoottext{b}{\cite{Anderson2018}.}
}
\end{threeparttable}      
\end{table}

   \begin{figure}
   \centering
   \includegraphics[width=0.48\textwidth]{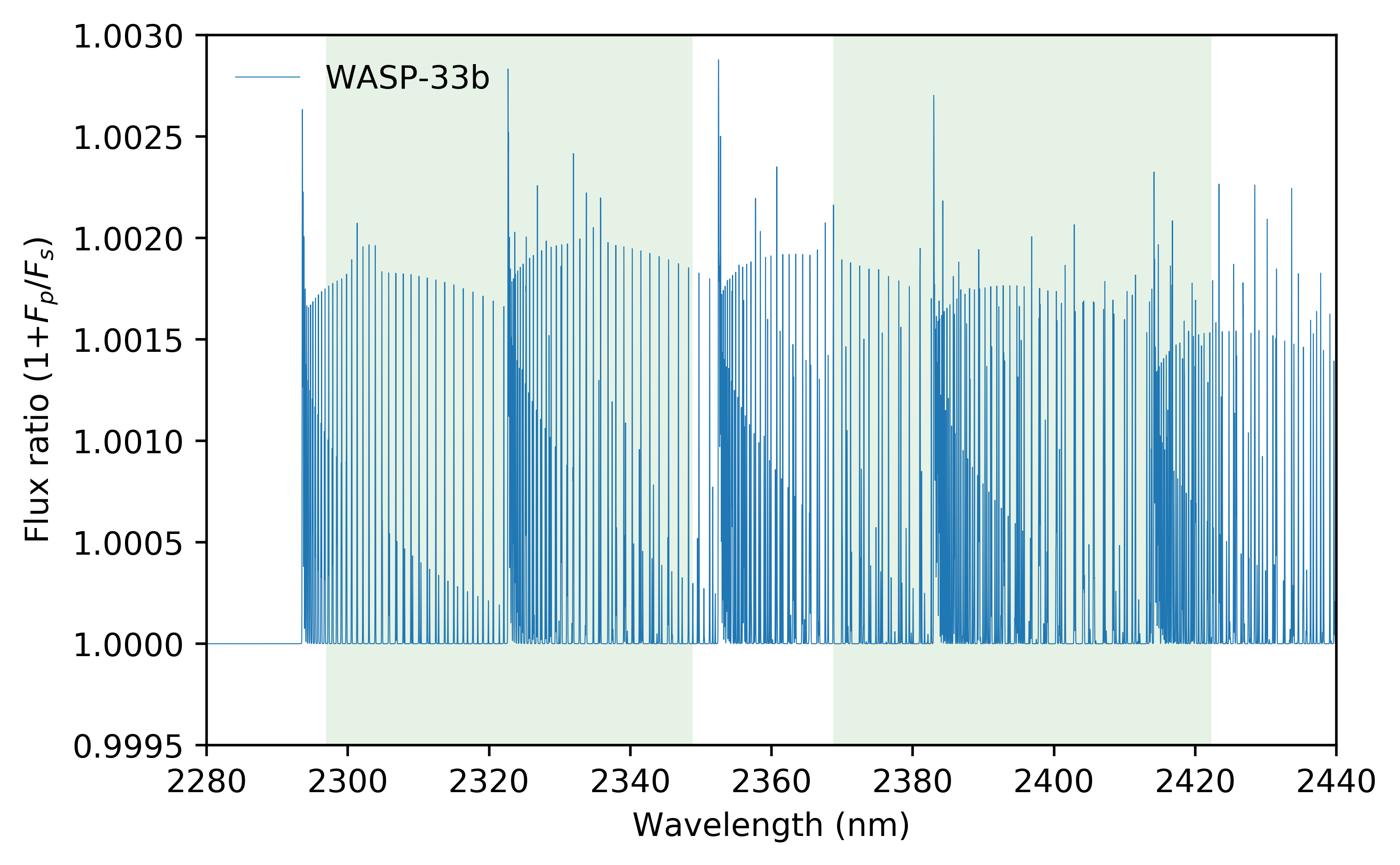}
   \includegraphics[width=0.48\textwidth]{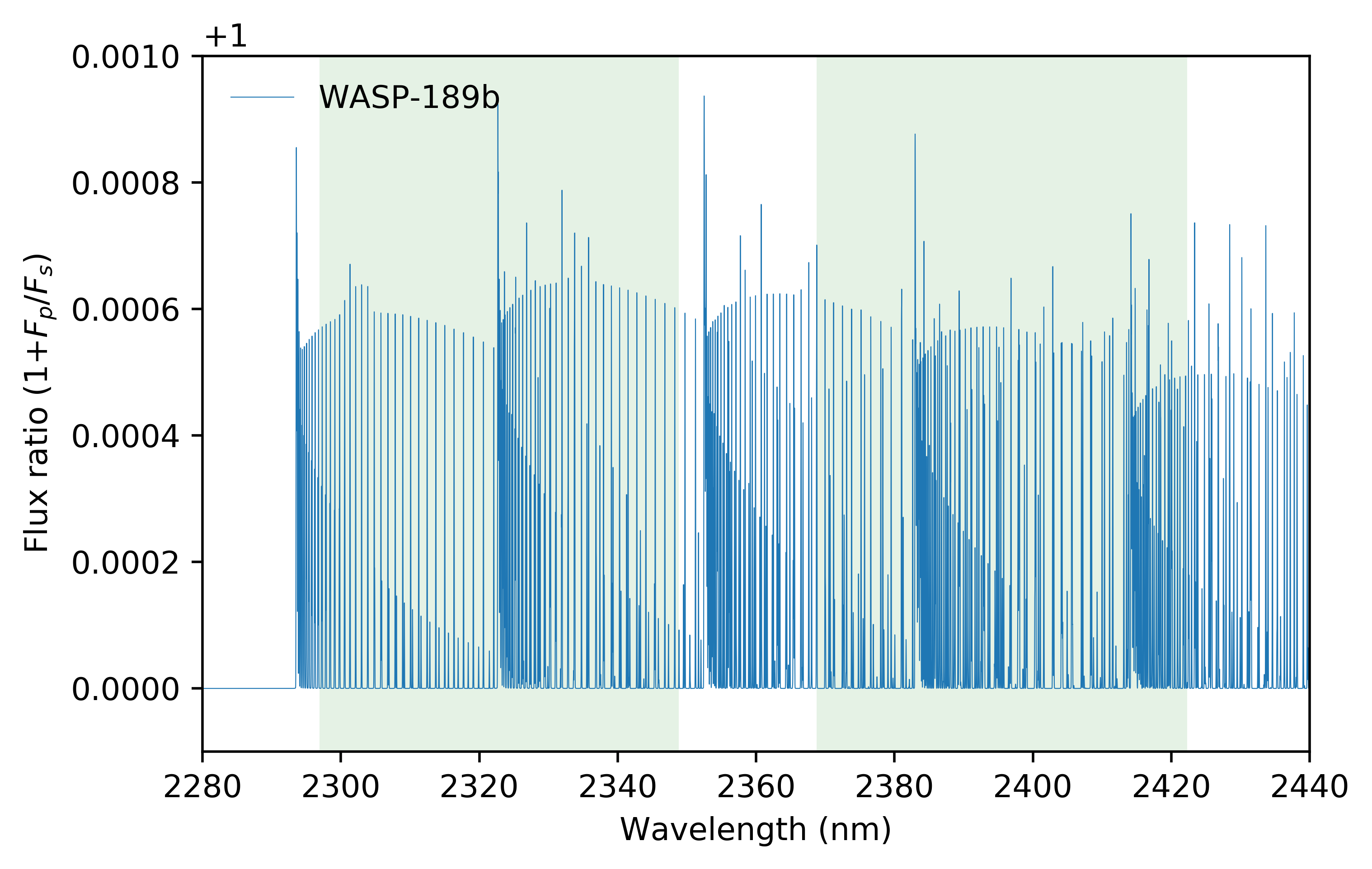}
      \caption{Modeled thermal emission spectra of CO lines. The green shadow indicates the wavelength coverage of the two spectral orders of the GIANO-B spectrograph. The spectra have been normalized and convolved to the instrument resolution.}
         \label{template}
   \end{figure}

   \begin{figure}
   \centering
   \includegraphics[width=0.45\textwidth]{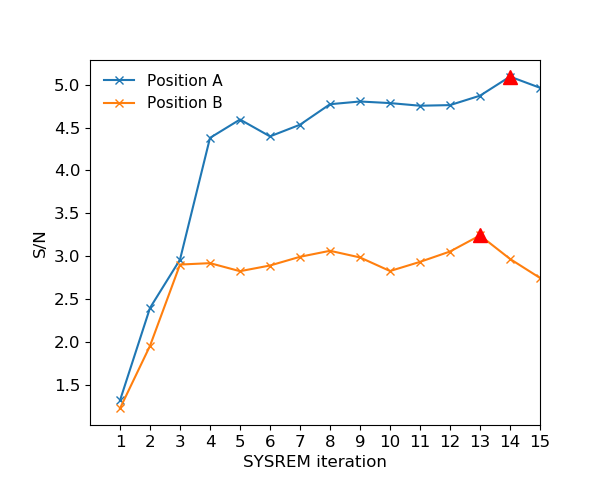}
   \includegraphics[width=0.45\textwidth]{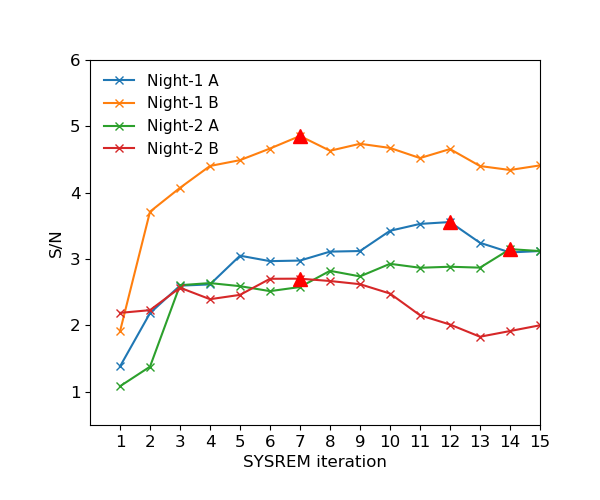}
      \caption{Detection significance with different \texttt{SYSREM} iteration numbers for WASP-33b (upper panel) and WASP-189b (lower panel). The red triangles denote the peak S/N values.}
         \label{App-sysrem}
   \end{figure}

\end{appendix}

\end{document}